\documentclass[]{aa}
\usepackage[varg]{txfonts}

\bibpunct{(}{)}{;}{a}{}{,}

\usepackage{hyperref}
\usepackage{graphicx}
\usepackage{booktabs}
\usepackage{multirow}
\usepackage{enumitem}
\usepackage{color}

\begin{document}

\defcitealias{zocchi:12}{ZBV12}

\title{A class of spherical, truncated, anisotropic models for application to globular clusters}
\author{Ruggero de Vita\inst{1}
\and Giuseppe Bertin\inst{1}
\and Alice Zocchi\inst{2}}
\institute{Università degli Studi di Milano, Dipartimento di Fisica,
via Celoria 16, Milano 20133, Italy
\and University of Surrey, Department of Physics,  
Guildford GU2 7XH, UK}
\date{Accepted for publication in Astronomy \& Astrophysics}

\abstract{Recently, a class of non-truncated radially-anisotropic models (the so-called $f^{(\nu)}$-models), originally constructed in the context of violent relaxation and modeling of elliptical galaxies, has been found to possess interesting qualities in relation to observed and simulated globular clusters. In view of new applications to globular clusters, we improve this class of models along two directions. To make them more suitable for the description of small stellar systems hosted by galaxies, we introduce a ``tidal” truncation (by means of  a procedure that guarantees full continuity of the distribution function). The new $f_T^{(\nu)}$-models are shown to provide a better fit to the observed photometric and spectroscopic profiles for a sample of 13 globular clusters studied earlier by means of non-truncated models; interestingly, the best-fit models also perform better with respect to the radial-orbit instability. Then we design a flexible but simple two-component family of truncated models, to study the separate issues of mass segregation and of multiple populations. We do not aim at a fully realistic description of globular clusters, to compete with the description currently obtained by means of dedicated simulations. The goal here is to try to identify the simplest models, that is, those with the smallest number of free parameters, but still able to provide a reasonable description for clusters that are evidently beyond the reach of one-component models: with this tool we aim at identifying the key factors that characterize mass segregation or the presence of multiple populations. To reduce the relevant parameter space, we formulate a few physical arguments (based on recent observations and simulations). A first application to two well-studied globular clusters is briefly described and discussed.}
\keywords{globular clusters: general  - Stars: kinematics and dynamics - globular clusters: individual: NGC 104  (47 Tuc), NGC 5139 (omega Cen)}
\maketitle


\section{Introduction}
\label{intro}
As a zeroth-order dynamical description, a class of models \citep{king:66} has
long and successfully been applied to globular clusters \citep{mclaughlin:05,carballo-bello:12,miocchi:13}. Standard spherical King models are meant to
describe round, nonrotating stellar systems made of a single stellar population,
for which the role of internal two-body relaxation has had time to act, bringing
the system close to a quasi-Maxwellian, isotropic distribution function; a
truncation is considered, to mimic the presence of tidal effects. The success of
the King models is largely based on their ability to fit the observed
photometric profiles (but see \citealt{mclaughlin:05} for a photometric test in
favor of models characterized by a milder truncation); the models are then used
to infer the general internal kinematical structure of globular clusters, which
is largely beyond the reach of direct observational tests. In recent years, with
the advent of high-resolution space and ground-based observations, the great
progress made in the acquisition of detailed information on the line-of-sight
and proper motion kinematics of these stellar systems has prompted a demand for
more complex dynamical models. In particular, many galactic globular clusters
are known to be characterized by significant rotation \citep{bellazzini:12,bianchini:13} and/or pressure anisotropy \citep{watkins:15}. Often, clusters
that are known to be characterized by longer relaxation times turn out to be
more anisotropic (see for example \citealt{zocchi:12}, hereafter
\citetalias{zocchi:12}, and \citealt{watkins:15}).

Regardless of their success, King models exhibit several internal
inconsistencies. The models are meant to describe tidally truncated stellar
systems, but in their original form they are spherical, in spite of the
stretching that tides are expected to impose. The models are chosen to reflect
the conditions of a collisionally relaxed state, but actually, outside their
half-mass radius, globular clusters and the models themselves are associated
with very long relaxation times \citep{harris:10}. These models are generally
applied as one-component models, that is, they are suited to describe stellar
systems made of a single homogeneous stellar population, yet, if collisional
relaxation is at work, it should generate significant mass segregation, with
heavier stars characterized by a distribution more concentrated than that of
lighter stars \citep{spitzer:69}.

Physically motivated models able to resolve some of the above-noted
inconsistencies, in relation to the shape and rotation of globular clusters,
have been constructed (in particular, see \citealt{heggie:95},
\citealt{bertin:08}, \citealt{varri:12}). As to the possible presence of
pressure anisotropy, for the case of nonrotating clusters, so far most studies
have resorted to the so-called Michie-King models \citep{michie:63}, which
introduce significant radial pressure in the outer parts by multiplication of
the underlying distribution function by a suitable angular-momentum-dependent
factor (see also the models recently proposed by \citealt{gieles:15}). In this
general picture, we might then consider models, such as those known as the
$f^{(\nu)}$ models, developed to represent the final state of collisionless
collapse under incomplete violent relaxation and successfully applied to the
study of bright elliptical galaxies \cite[e.g., see][]{trenti:05}. Even though
it remains to be proved that the formation of globular clusters, or at least of
some globular clusters, follows this route, some recent investigations have
looked into this possibility.

A general trend in the direction of radial pressure in the outer regions has
been noted also in recent simulations of the evolution of globular clusters
\citep{tiongco:16}. [Eventually, if external tidal fields are present, the
outermost regions of the cluster may be characterized by isotropy or mild
tangential anisotropy, as also suggested by \citet{vesperini:14}.] In a recent
paper (\citetalias{zocchi:12}), the class of spherical $f^{(\nu)}$ models has
been used to study a sample of Galactic globular clusters under different
relaxation conditions and compared to the performance of the standard spherical
King models. This exploratory investigation indicates that for some clusters the
use of $f^{(\nu)}$ models is encouraged, although, being non-truncated, these
models are obviously at a disadvantage in describing the outer parts of the
available photometric profiles. In addition, some of the best-fit radially
anisotropic models thus identified actually turn out to be too anisotropic, so
that they might be prone to the radial-orbit instability (and thus not
acceptable for interpreting the observations). The first goal of the present
paper is to introduce a truncation to the $f^{(\nu)}$ models and to test whether
this new class of models is capable of a more satisfactory fit to the sample of
globular clusters studied by \citetalias{zocchi:12}.

The second objective of the paper is to extend the newly constructed
$f_T^{(\nu)}$ models to the case of two-component systems. For globular
clusters, there are at least two important reasons to address more complex
models of this kind.

One of the main effects related to collisionality is that of mass segregation.
Thus a more realistic dynamical framework for the modeling of globular cluster
has been sought in terms of multi-component models  \cite[e.g., see][]
{dacosta:76,gunn:79,merritt:81,miocchi:06}, which basically represent an extension of the standard King (or Michie-King) models. Naively (i.e., in the normal context
of kinetic systems), we would expect collisions to enforce a sort of
equipartition, in which the velocity dispersion $\sigma$ of stars of mass $m$
should scale as $\sigma \sim m^{-1/2}$. The process is complicated by the global
and inhomogeneous nature of self-gravitating systems. It has also been argued
that in the core of globular clusters complete equipartition cannot be achieved
as a result of the so-called Spitzer instability. In particular,
\citet{spitzer:69} suggested that, in a two-component system in virial
equilibrium, the condition of equipartition in the core is precluded if the
total mass of the heavy stars exceeds a certain fraction of the total mass of
the cluster. Spitzer's criterion has been extended by \citet{vishniac:78} to
cover systems with a continuous distribution of masses. These theoretical
arguments have been revisited by means of recent simulations \cite[see][]
{trenti:13}, in which only partial equipartition is ``observed'' to follow from
the cumulative action of star-star collisions. In any case, a certain degree of
mass segregation appears to emerge from the observations of several globular
clusters \cite[see][]{anderson:10,goldsbury:13,dicecco:13,bellini:14}.

A second, physically separate reason to address the issue of two-component
models is given by the relatively recent finding that globular clusters host
multiple stellar populations. In many observed cases, the suggested
interpretation is that clusters have been the site of multiple generations of
stars \cite[see][]{lardo:11,gratton:12}, so that the stars can be divided into the groups of
the first and the second generation, and these groups may be associated with
different dynamical properties, such as concentration or degree of anisotropy \cite[see][]{richer:13,bellini:15}.

For the second goal of the paper, that is, the construction of two-component
models of the $f_T^{(\nu)}$ form, to keep the number of free parameters low, we
formulate some physical hypotheses (based on observations and/or simulations)
that correspond to the picture of mass segregation. A comparison with observed
cases should be able to support or disprove the physical assumptions made in the
modeling procedure. Our approach is complementary to that of constructing multiparameter
models as diagnostic tools (see \citealt{dacosta:76},
\citealt{gunn:79}, \citealt{gieles:15}).

The paper is organized as follows. In Sect.~\ref{sec:2} we introduce and
construct the new class of truncated anisotropic $f_T^{(\nu)}$ models. In
Sect.~\ref{sec:3} we extend it to the two-component case. In Sect.~\ref{sec:4}
we apply the one-component models to fit a sample of 13 galactic globular
clusters. For NGC 5139 ($\omega$ Cen) and NGC 104 (47 Tuc), we also present the
results of the fits performed by means of two-component $f_T^{(\nu)}$ models.
Finally, in Sect.~\ref{sec:5}, we draw our conclusions.

\section{One-component models}
\label{sec:2}
Studies of the dynamics of elliptical galaxies have investigated the picture of galaxy formation by incomplete violent relaxation from collisionless collapse. There are ways to translate this picture into an appropriate choice of the relevant distribution function to represent the current state of ellipticals. The choice is not unique and various options have been explored. One particular choice reflects a conjecture on the statistical foundation of the relevant distribution function \cite[see][]{stiavelli:87}. This is a family of partially relaxed models. The models are called $f^{(\nu)}$ models and their properties have been studied extensively in more recent papers \cite[see][]{bertin:03,trenti2:05}. They are based on the following distribution function
\begin{equation}
\label{fnuDF}
f^{(\nu)} = A \exp{\left[-a E - d\left(\frac{J^2}{|E|^{3/2}}\right)^{\nu/2} \right]},
\end{equation}
where $A$, $a$, and $d$ are positive constants. For applications, a given value of $\nu\approx1$ is usually taken as a fixed parameter. Here $E=v^2/2+\Phi(r)<0$ and $J=|\bold{r}\times\bold{v}|$ represent the specific energy and the magnitude of the specific angular momentum of a single star subject to a spherically symmetric mean potential $\Phi(r)$. The self-consistent models based on this distribution function define a family of anisotropic, non-truncated models. The following subsections are devoted to the formulation of a truncated distribution function as a generalization of Eq.~\eqref{fnuDF} and to the analysis of the main dynamical properties found for the resulting new classes of anisotropic truncated models. 

\subsection{Truncation}
\label{subsec:2.1}
As also noted by \citet{davoust:77}, the truncation prescription is not unique (the structural properties associated with different types of truncation are described by \citealt{hunter:77}); indeed, the distribution functions considered in that article differ from one another for the smoothness of their energy gradients in correspondence with the energy cut-off. In this respect, we decided to proceed to the truncation of the $f^{(\nu)}$ models with $\nu=1$ in the following way. The distribution function
\begin{equation}
\label{fnuTDF}
f_T^{(\nu)} =
\begin{cases}
A \exp{\left[-a(E-E_e)- \frac{dJ}{|E-E_e|^{\frac{3}{4}}} \right]}		&\text{if $E<E_e$}\\
0												&\text{if $E\ge E_e$}
\end{cases}
\end{equation}
for $J\neq0$, vanishes at the cut-off energy $E_e$ together with all its derivatives (the quantities $A, E_e, a$, and $d$ are constants).
The two-parameter family of one-component models is then constructed by solving the Poisson equation:
\begin{equation} \label{PE}
\nabla^2 \Phi (r)= 4\pi G \int f_T^{(\nu)} \ d^3v
\end{equation}
for the gravitational potential $\Phi(r)$. In our case, the distribution function is anisotropic, so that the density on the right-hand side of Eq.~\eqref{PE} can be reduced to a two-dimensional integral, which depends on radius explicitly and implicitly, through the unknown $\Phi(r)$. Thus, if we define dimensionless quantities such as the potential $\psi=-a(\Phi-E_e)$, the radius $\xi=a^{1/4}d r$, and the velocity $\omega^2=(a/2)v^2$, the integral is proportional to
\begin{equation}
\label{hat_rho}
\hat{\rho}(\xi,\psi)=\int_0^\pi\int_0^{\sqrt{\psi}}\hat{f}_T^{(\nu)}(\xi,\psi,\omega,\zeta)\omega^2\sin\zeta d\zeta d \omega~,
\end{equation}
where 
\begin{equation}
\label{fnuT_om_zeta}
\hat{f}_T^{(\nu)}(\xi,\psi,\omega,\zeta)=4\sqrt{2}\pi\exp{\left[-\omega^2+\psi-\frac{\sqrt{2}\xi\omega\sin\zeta}{|\omega^2-\psi |^{3/4}}\right]}~,
\end{equation}
and $\zeta$ is the angle between the position $\bold{r}$ and the velocity $\bold{v}$ of a single star.
The resulting dimensionless form of Eq.~\eqref{PE} is given by 
\begin{equation}
\label{PEfnuT}
\frac{d^2}{d\xi^2}\psi + \frac{2}{\xi} \frac{d}{d\xi}\psi = -\frac{1}{\gamma} \hat{\rho} (\xi,\psi)~,
\end{equation}
where we have introduced the dimensionless parameter $\gamma=ad^2/(4\pi GA)$. This differential equation is integrated under the boundary conditions $\psi(0)=\Psi$ and $(d\psi/d\xi)(0)=0$ out to the truncation radius $\xi_{tr}$, where the dimensionless potential vanishes. Hence, the self-consistent problem for the dimensionless potential reduces to a family of second-order differential equations defined by two structural parameters: the central dimensionless potential $\Psi$ and $\gamma$. 
We have performed the integration of Eq.~\eqref{PEfnuT} with an adaptive fourth-order Runge-Kutta method. At every integration step, the two-dimensional integral on the right-hand side has been computed by means of the Chure routine in the C-package Cuba \cite[see][]{hahn:05}.

\subsection{The parameter space}
\label{subsec:2.2}
The non-truncated models are characterized by a specific relation between the parameters $\Psi$ and $\gamma$ (see the plot of $\gamma(\Psi)$ in Fig.~1 of \citealt{trenti:05}). In particular, for a given value of $\Psi$ the corresponding value of $\gamma$ is fixed by the requirement of a Keplerian decay of the gravitational potential ($\Phi\sim-1/r$) at large radii.
For the models with  $\nu=1$, in the range $0\lesssim\Psi\lesssim15$, the function $\gamma(\Psi)$ presents a pronounced peak at $\Psi \approx 5.5$; for higher values of $\Psi$, $\gamma$ decreases, reaches about half of its peak value at $\Psi \approx 10$, and then stays approximately constant.

\begin{figure} 
\resizebox{\hsize}{!}{\includegraphics{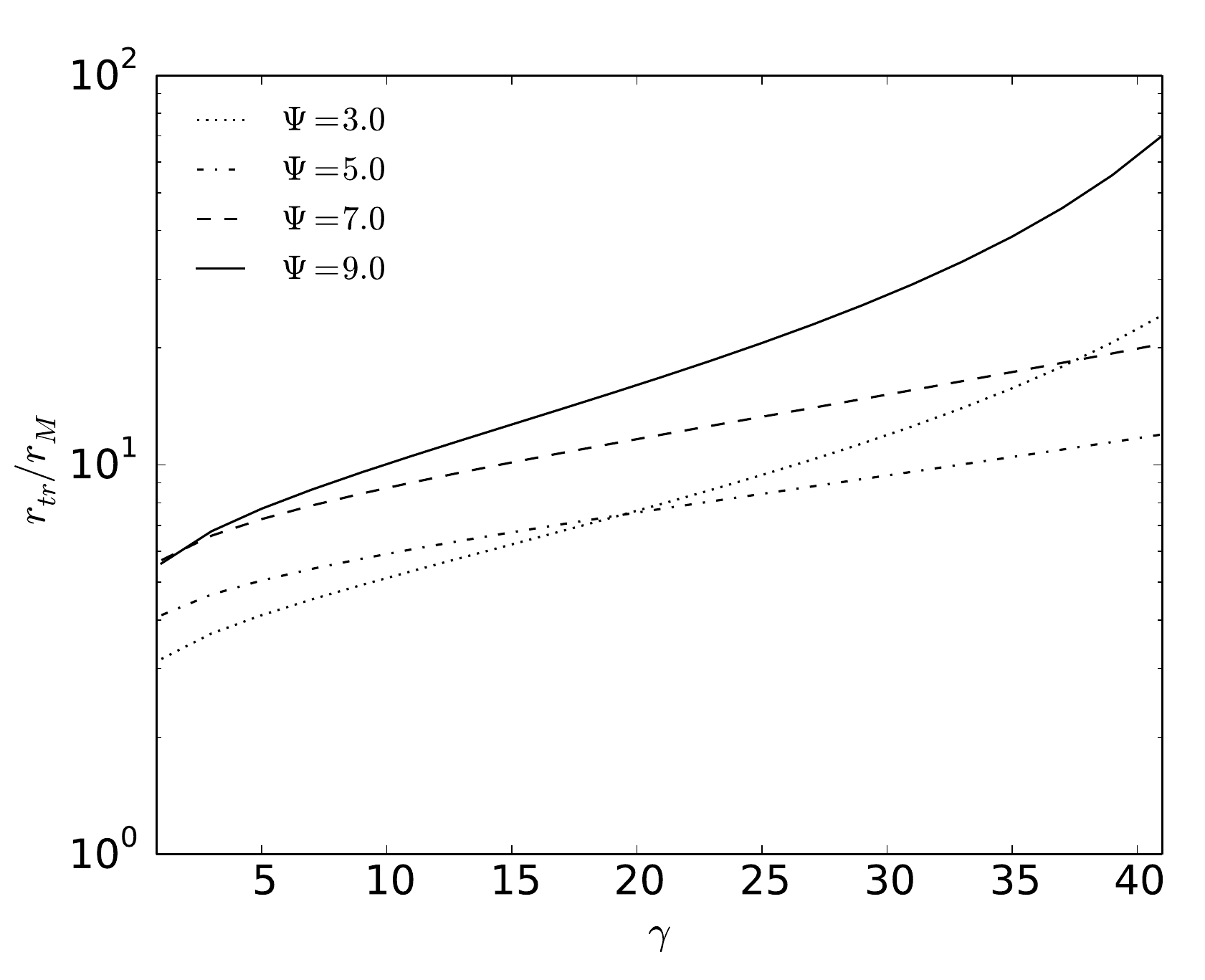}} 
\caption{The quantity $r_{tr}/r_M$ is plotted as a function of $\gamma$, for selected values of $\Psi$.}
\label{fig:truncation}
\end{figure}

In our models $\gamma$ is left as a free parameter. However, since, for a given $\Psi$, there is a maximum value $\gamma_{max}$ beyond which the models do not present any truncation, the parameter space is confined to the region that is under the curve $\gamma(\Psi)$ found for the non-truncated models. For a given $\Psi$, the non-truncated models are recovered in the limit $\gamma\rightarrow\gamma_{max}$; indeed, as shown in Fig.~\ref{fig:truncation},  the ratio of the truncation radius $r_{tr}$ to the half-mass radius $r_M$ is an increasing function of $\gamma$.

\begin{figure} 
\resizebox{\hsize}{!}{\includegraphics{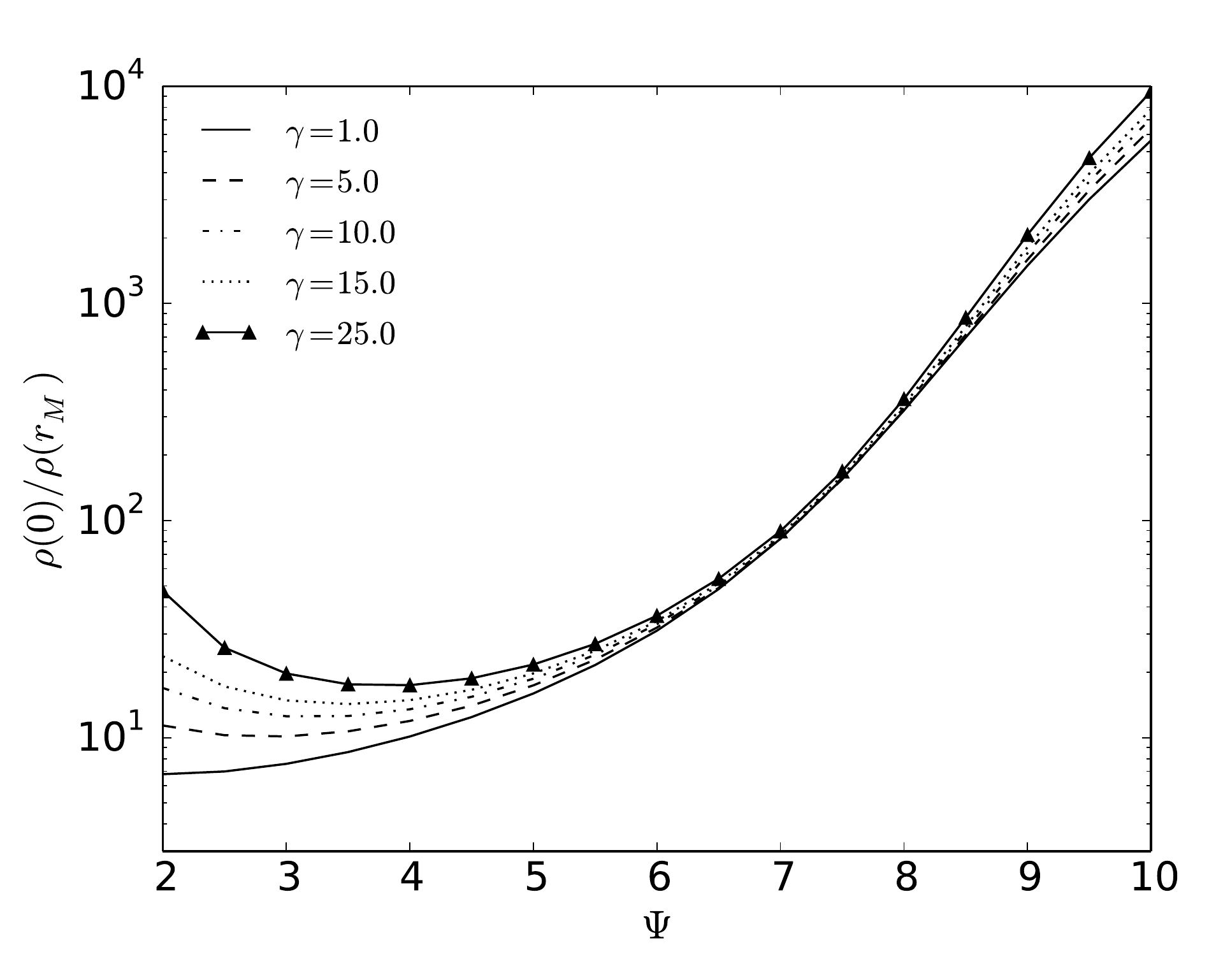}} 
\caption{The quantity $\rho(0)/\rho(r_M)$ is plotted as a function of $\Psi$, for selected values of $\gamma$.}
\label{fig:concentration}
\end{figure}

The parameter $\Psi$ is identified with the concentration of the model. Another measure of the central concentration is the ratio $\rho(0)/\rho(r_M)$ of the central density to the density calculated at the half-mass radius $r_M$. In Fig.~\ref{fig:concentration} we plot this quantity as a function of $\Psi$. We note that for high values of $\gamma$ the relation is non-monotonic. For $5.5\lesssim\Psi\lesssim8.5$ the relation is monotonic and characterized by a weak dependence on $\gamma$ .

\subsection{Intrinsic profiles}
\label{subsec:2.3}
All the radial profiles of physical interest can be derived by taking moments of the distribution function $f$.  If we consider the natural velocity coordinate system $(v_r,v_\theta,v_\varphi)$, the velocity dispersion tensor is diagonal with $\sigma_{\theta\theta}^2=\sigma_{\varphi\varphi}^2$. Explicitly, by defining a tangential component of the velocity dispersion tensor as $\sigma_T^2=\sigma_{\theta\theta}^2+\sigma_{\varphi\varphi}^2$, we have
\begin {align}
\sigma_{rr}^2			&=\frac{2}{a}\frac{1}{\hat{\rho}}\int_0^\pi\int_0^{\sqrt{\psi}}\hat{f}^{(\nu)}_{T}(\xi,\psi,\omega,\zeta)\omega^4\cos^2\zeta \sin\zeta d\zeta d\omega~,\label{sigr}\\
\sigma_T^2			&=\frac{2}{a}\frac{1}{\hat{\rho}}\int_0^\pi\int_0^{\sqrt{\psi}}\hat{f}^{(\nu)}_{T}(\xi,\psi,\omega,\zeta)\omega^4 \sin^3\zeta d\zeta d\omega~\label{sigT},
\end{align}
where we have used the definitions given in Eqs.~\eqref{hat_rho}-\eqref{fnuT_om_zeta} and the relations: $v_r^2=v^2\cos^2\zeta$ and $v_T^2=v^2_\theta+v_\varphi^2=v^2\sin^2\zeta$. 
For simplicity, in the following we will use the notation $\sigma_r^2=\sigma_{rr}^2$. Once the dimensionless potential profile is obtained by solving the Poisson equation, the velocity dispersion profiles can be calculated as two-dimensional integrals with the same procedure described in Subsect.~\ref{subsec:2.1}.
\begin{figure*}
 \centerline{
 \includegraphics[width=0.5\textwidth]{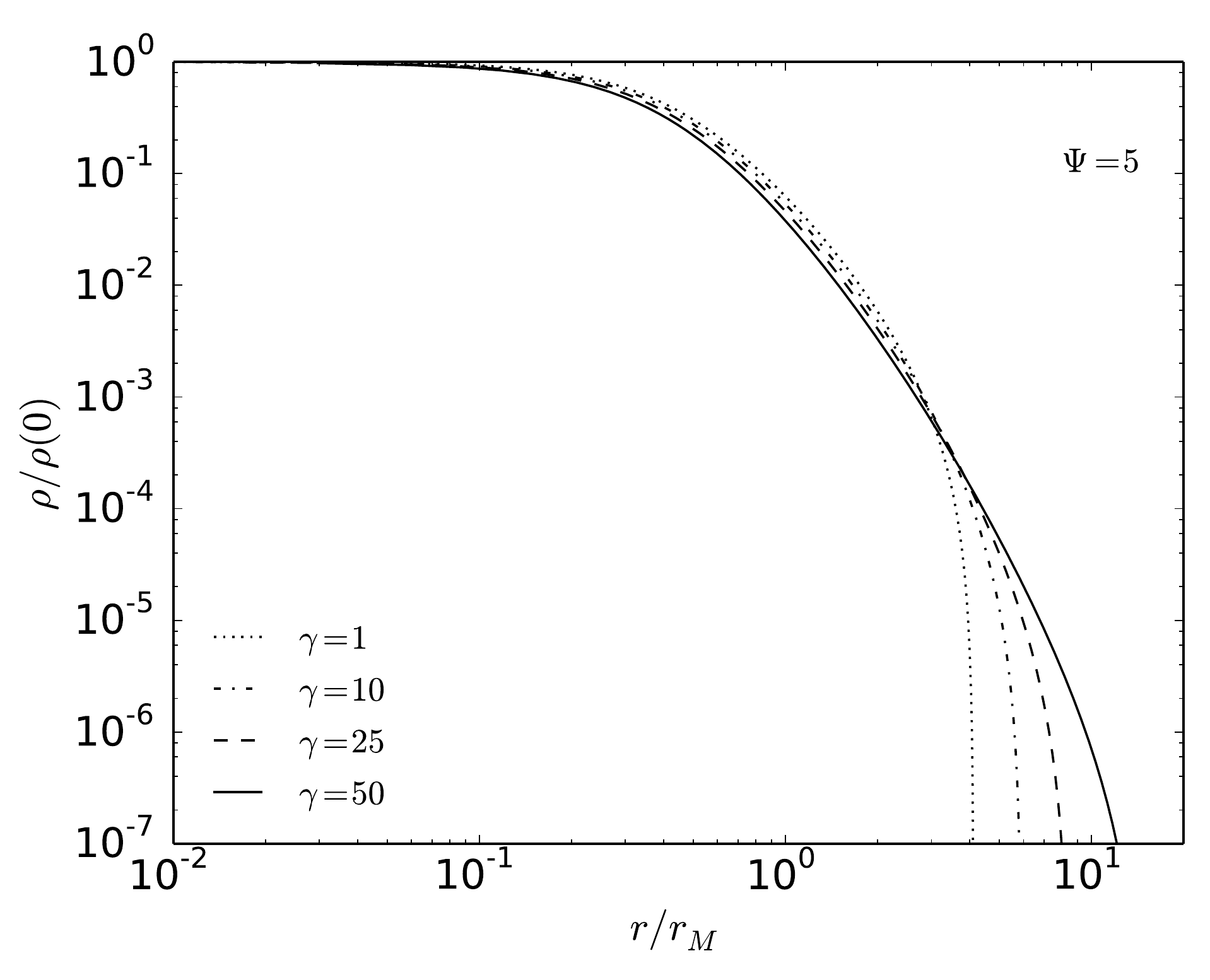}
 \includegraphics[width=0.5\textwidth]{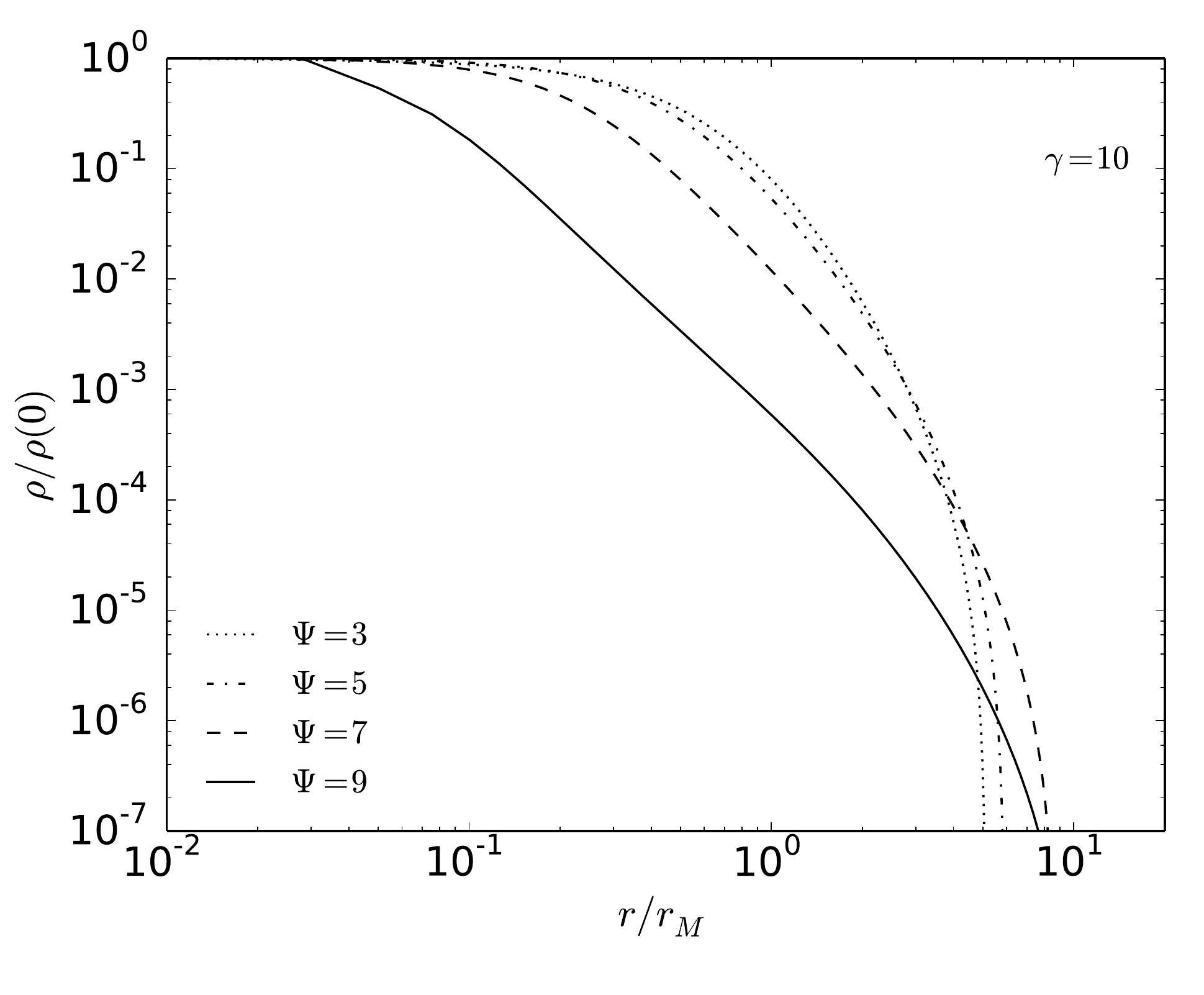}
 }
 \caption{The left frame shows the normalized density profile for selected values of $\gamma$ at fixed $\Psi$; the right frame shows the normalized density profile for selected values of $\Psi$ at fixed $\gamma$.}
 \label{fig:dens}
 \end{figure*}
 
\begin{figure*}
 \centerline{
 \includegraphics[width=0.5\textwidth]{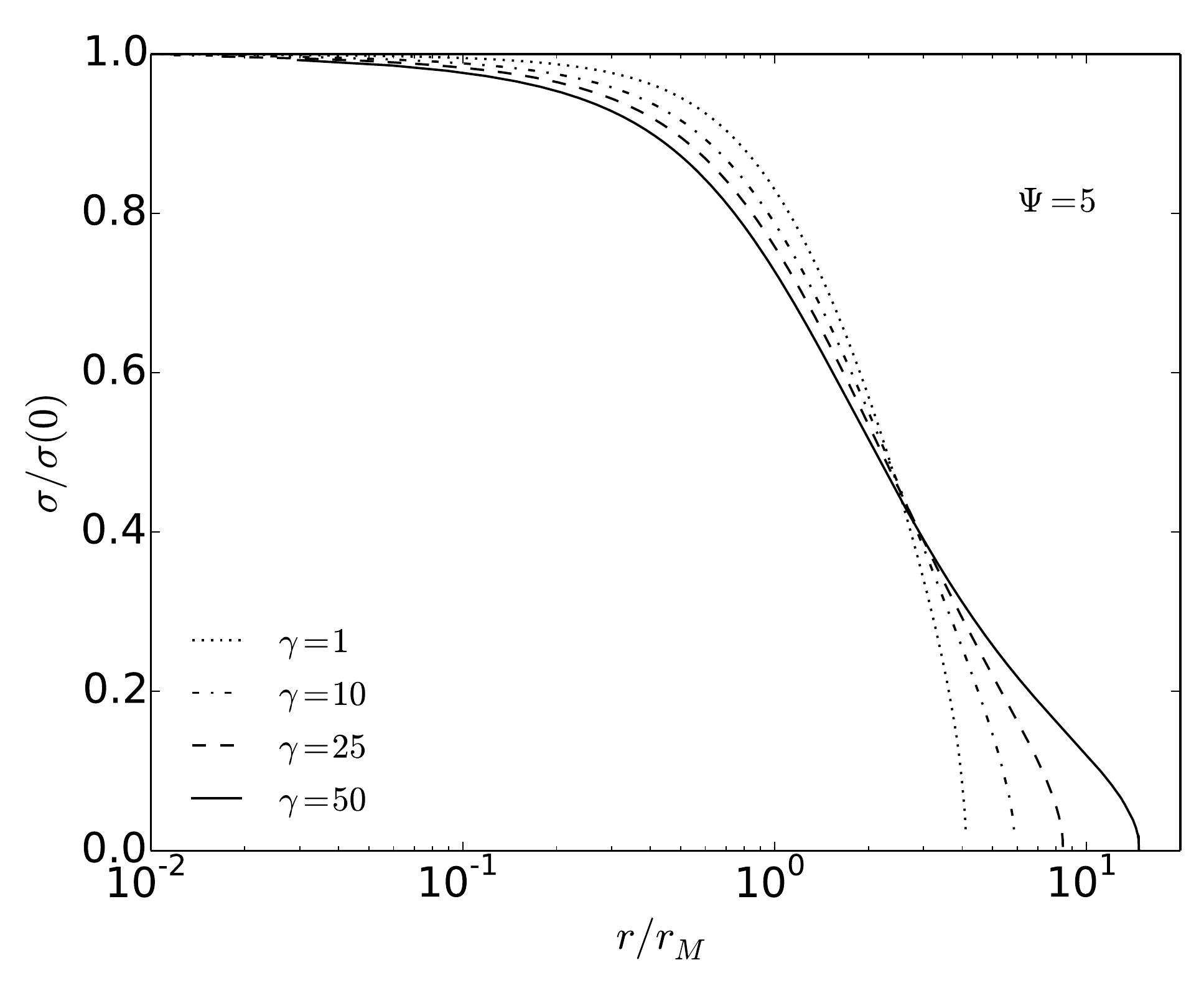}
 \includegraphics[width=0.5\textwidth]{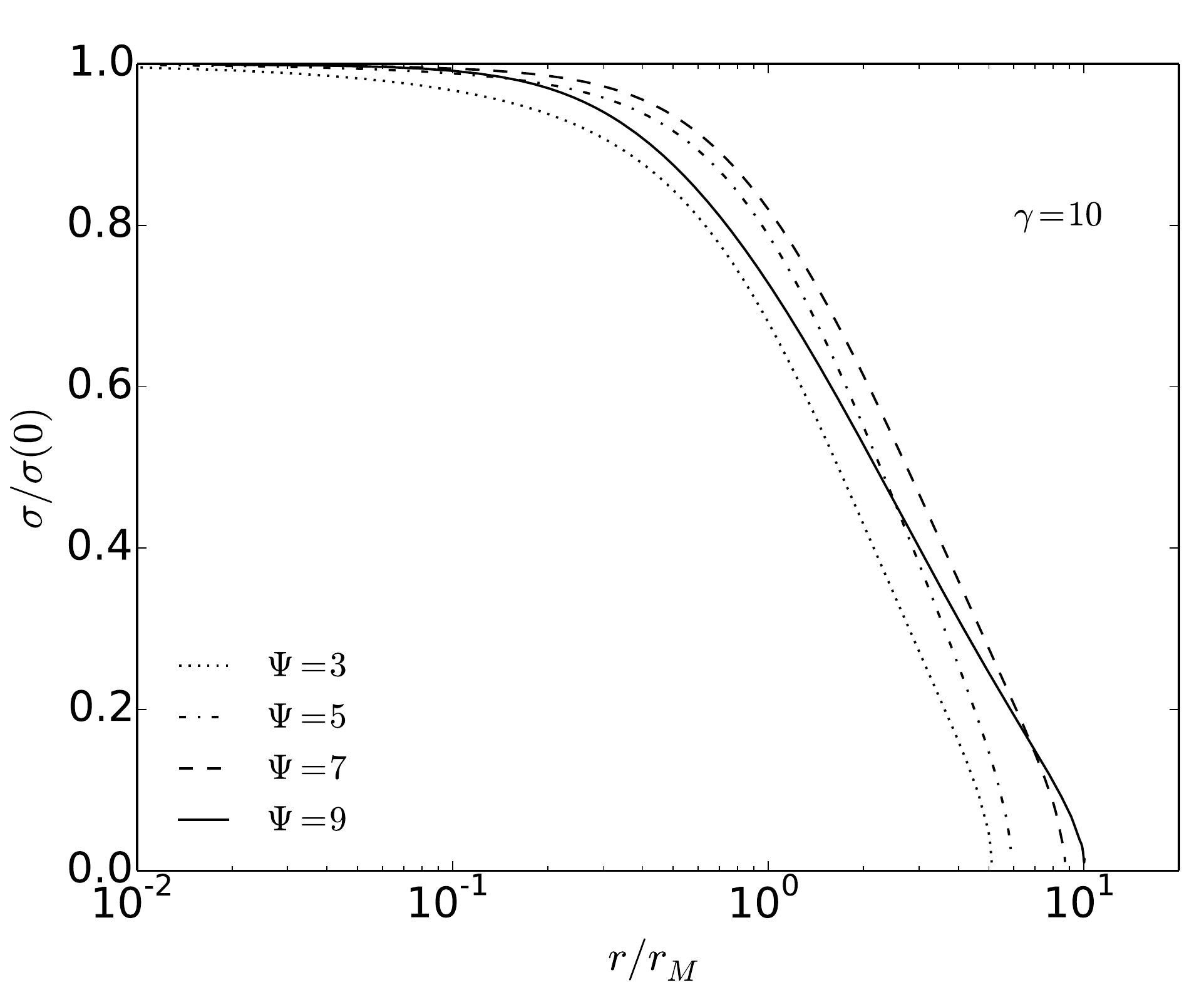}
 }
 \caption{The left frame shows the normalized total velocity dispersion profile for selected values of $\gamma$ at fixed $\Psi$; the right frame shows the normalized total velocity dispersion profile for selected values of $\Psi$ at fixed $\gamma$.}
 \label{fig:disp}
 \end{figure*}
 
For the one-component $f_T^{(\nu)}$ models, in Fig.~\ref{fig:dens} and Fig.~\ref{fig:disp}, we plot some intrinsic profiles of the density and the total velocity dispersion (defined by  $\sigma^2=\sigma_r^2+\sigma_T^2$).

\subsection{Anisotropy}
\label{subsec:2.4}
A local measure of the pressure anisotropy is given by the function $\alpha(r)=2-\sigma_T^2/\sigma_r^2$. 
\begin{figure*}
 \centerline{
 \includegraphics[width=0.5\textwidth]{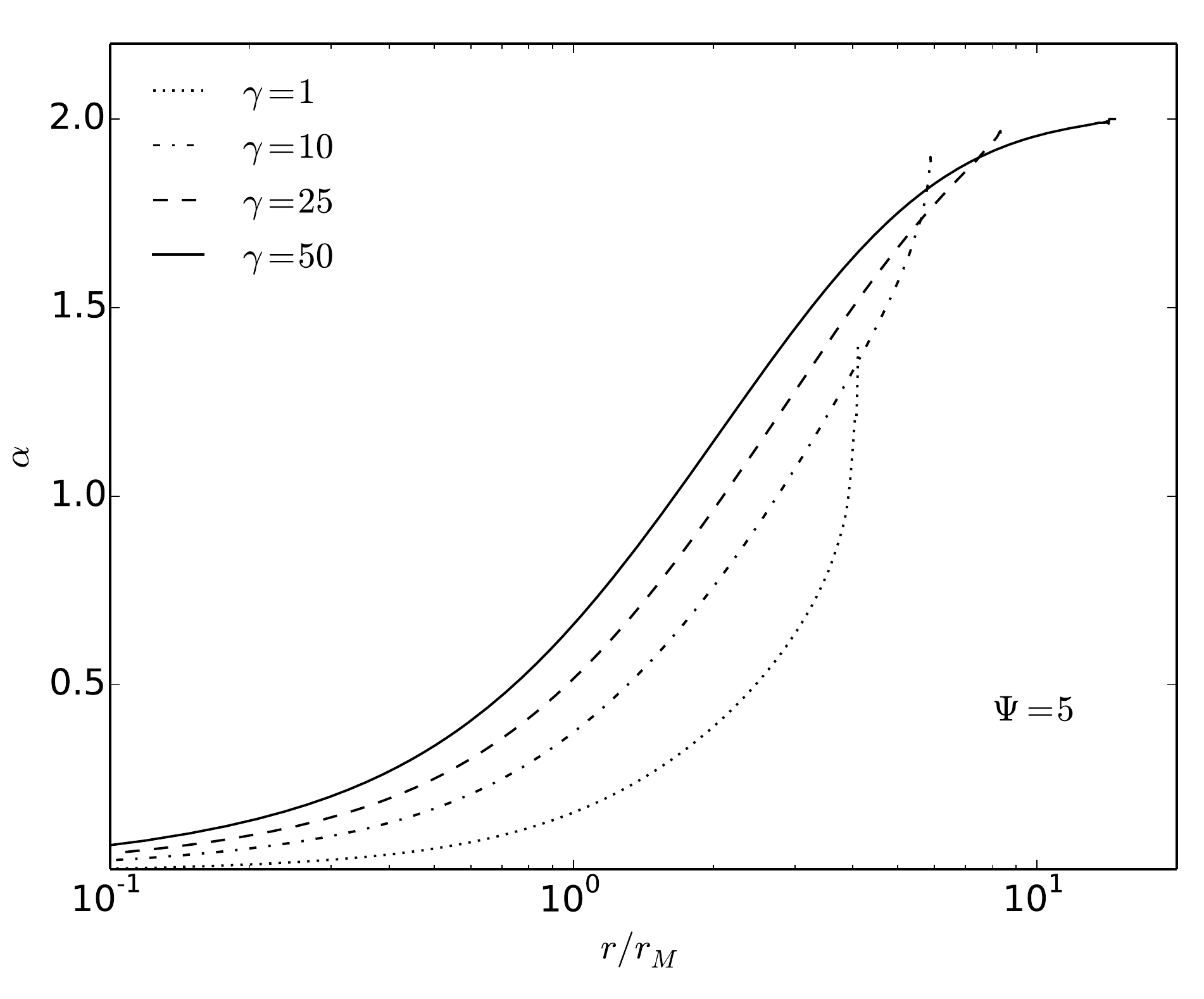}
 \includegraphics[width=0.5\textwidth]{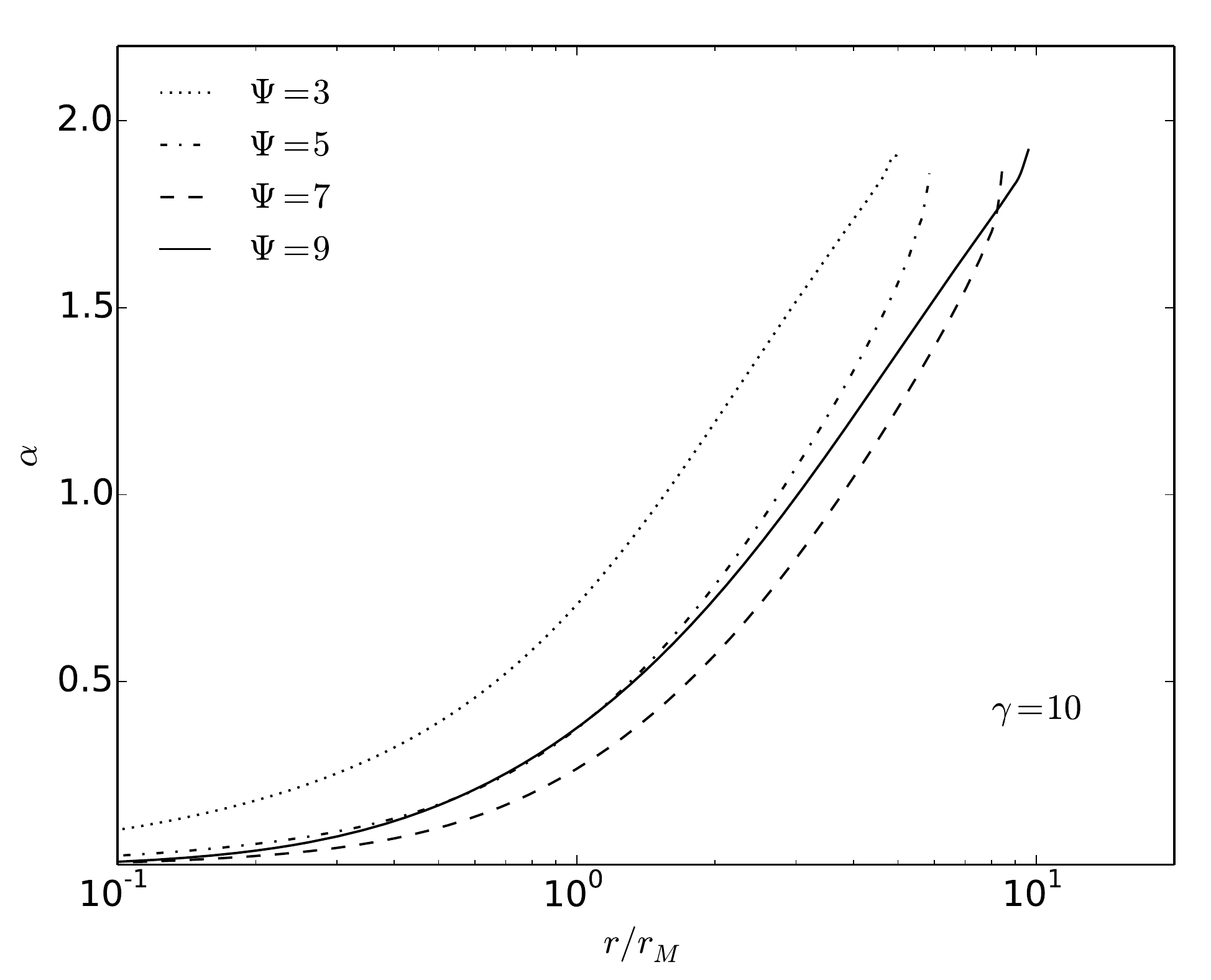}
 }
 \caption{ The left frame shows the anisotropy profile $\alpha(r)$ for selected values of $\gamma$ at fixed $\Psi$; the right frame shows the anisotropy profile for selected values of $\Psi$ at fixed $\gamma$. Where a curve terminates, the truncation radius is reached.}
 \label{fig:alpha}
 \end{figure*}
In Fig.~\ref{fig:alpha} we show some representative anisotropy profiles. The models are characterized by an isotropic core and a radially-biased anisotropic envelope. 

The radial extent of the isotropic core can be measured by means of the anisotropy radius $r_\alpha$ defined as the radius where $\alpha=1$. 
\begin{figure} 
\resizebox{\hsize}{!}{\includegraphics{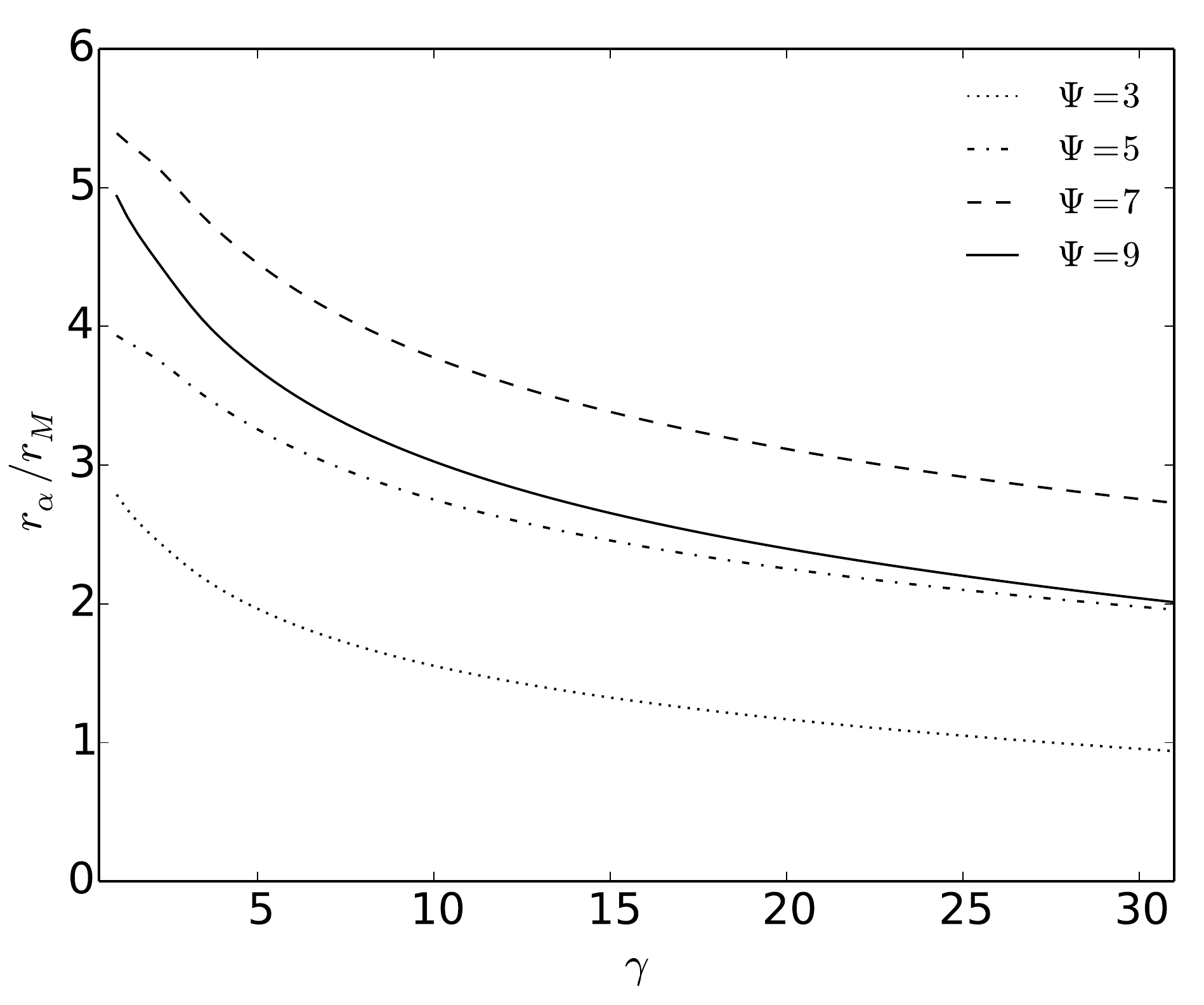}} 
\caption{The ratio of the anisotropy radius $r_\alpha$ to the half-mass radius $r_M$ as a function of $\gamma$ for selected values of $\Psi$. At given $\Psi$, models with smaller $\gamma$ are characterized by a more extended isotropic core.}
\label{fig:rarM}
\end{figure}
\begin{figure} 
\resizebox{\hsize}{!}{\includegraphics{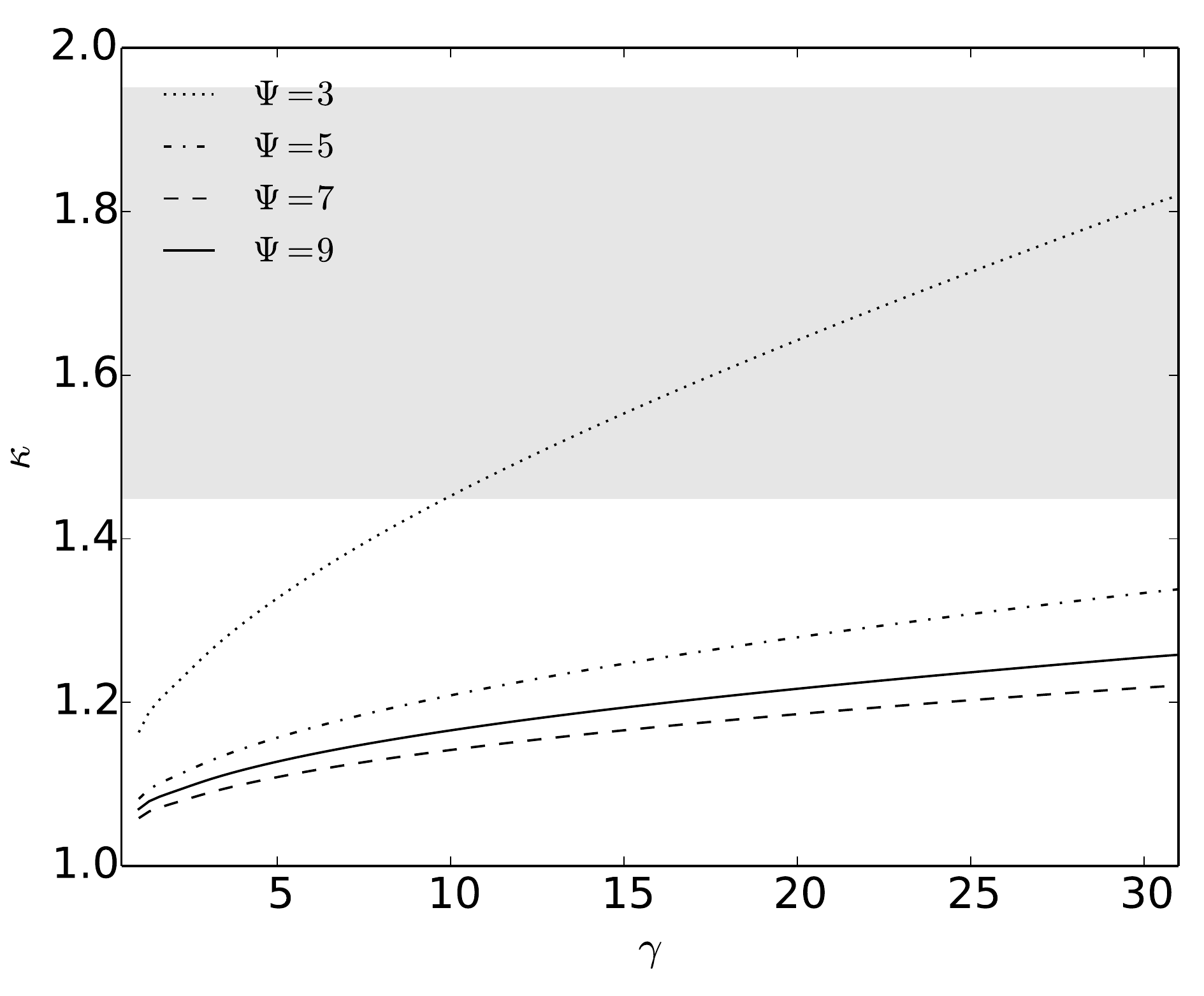}} 
\caption{Global anisotropy parameter $\kappa=2K_r/K_T$ for selected values of the parameter $\Psi$. The grey area indicates the region of the threshold for the onset of the radial orbit instability.}
\label{fig:kappa}
\end{figure} 
The ratio $r_\alpha/r_M$ of the anisotropy radius to the half-mass radius as a function of $\gamma$ is shown in Fig.~\ref{fig:rarM}. At fixed $\Psi$, models with higher $\gamma$ are characterized by lower values of $r_\alpha/r_M$. 

This trend is confirmed by the behavior of the ratio $\kappa=2K_r/K_T$ of twice the total radial kinetic energy $K_r$ to the total tangential kinetic energy $K_T$, which is often used to measure the degree of global anisotropy of the system. This parameter is related to a well-known criterion for the onset of the radial-orbit instability \citep{polyachenko:81}: instability occurs if $\kappa$ exceeds a model-dependent threshold, $\kappa\gtrsim1.7\pm0.25$. Figure \ref{fig:kappa} shows the monotonic increasing dependence of $\kappa$ on $\gamma$. Therefore, truncated models are generally more isotropic than the corresponding non-truncated models.

\subsection{Virial coefficient}
\label{subsec:2.5}
The virial coefficient (for more details see \citealt{bertin:02}) can be defined as
\begin{equation}
K_V=\frac{G\Upsilon_*L}{R_e\sigma_0^2},
\end{equation}
where $\sigma_0$ is the ``central'' velocity dispersion,\footnote{In the following we will consider $\sigma_0$ as the mean value of the line-of-sight velocity dispersion on the cylindrical volume with projected radius $R_e/8$ and length $2r_{tr}$.} $\Upsilon_*$ is the stellar mass-to-light ratio in the band used for the determination of the luminosity $L$, and the effective radius $R_e$ is the projected radius of the disk containing half of the total luminosity of the cluster.

Once the best-fit model for a given cluster is found from the photometric fit, the virial coefficient can be calculated, and thus used in order to infer the total dynamical mass from a measurement of $\sigma_0$ (under the hypothesis of a single stellar component). This procedure is very useful, particularly for those cases in which the kinematic profiles are poor or affected by large uncertainties. 

\begin{figure} 
\resizebox{\hsize}{!}{\includegraphics{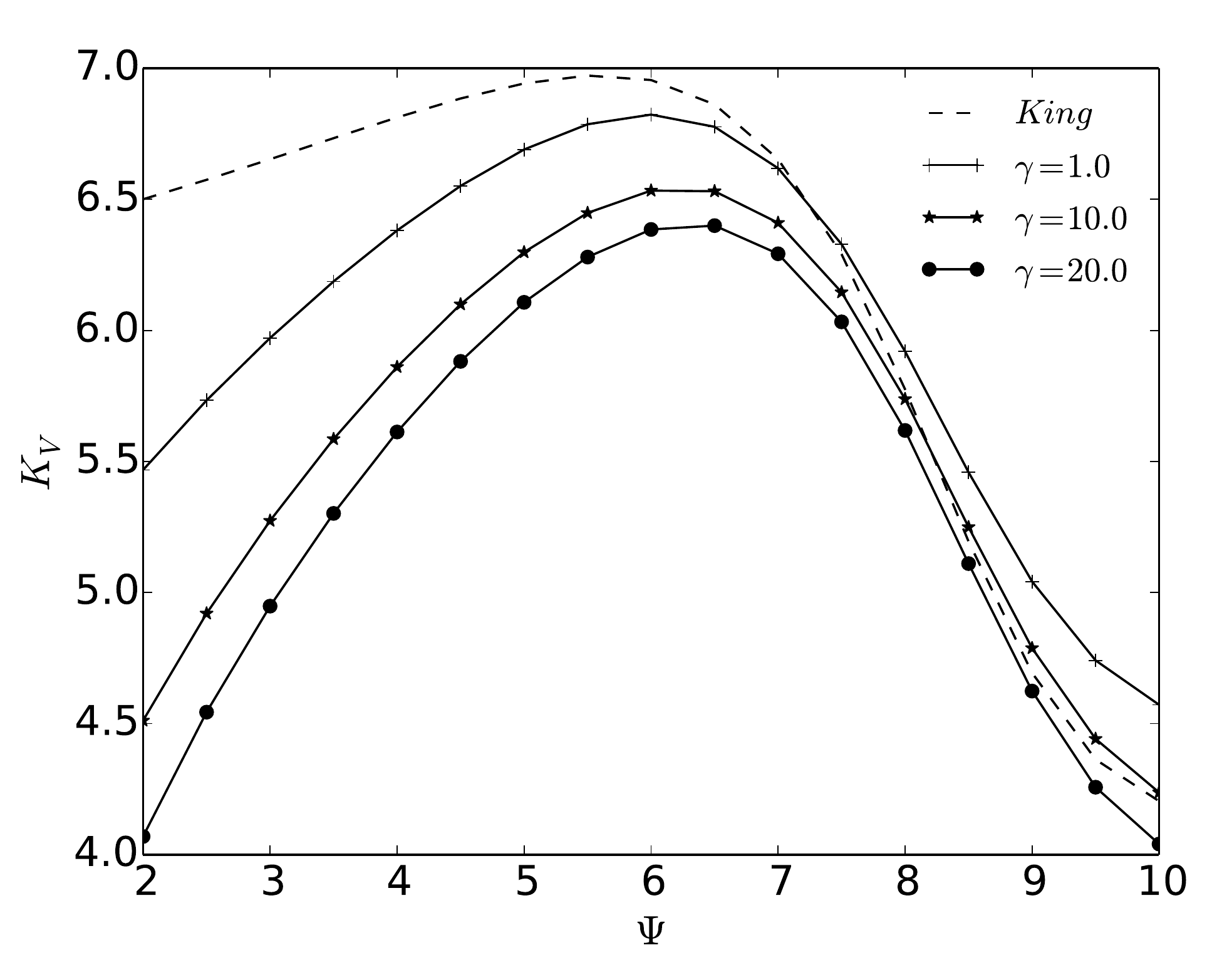}} 
\caption{Virial coefficient $K_V$ for selected values of $\gamma$ and for the King models.}
\label{fig:K_V}
\end{figure}

In Fig.~\ref{fig:K_V} we show the value of $K_V$  as a function of the central dimensionless potential $\Psi$ for selected values of $\gamma$ and for the King models. The difference between the various curves can be significant, particularly for low values of $\Psi$.

\section{Two-component models}
\label{sec:3}
Starting from the truncated models described in the previous subsections, we introduce the two distribution functions:
\begin{equation}
\label{fnu2T}
f^{(\nu)}_{T,i}(E,J)= 	\begin{cases}
				A_i\exp{\left[-a_i(E-E_i)-d_i\frac{J}{\left| E-E_i\right|^{3/4}} \right]} 	&\text{if $E<E_i$} \\
				0													&\text{if $E\ge E_i$.}
				\end{cases}
\end{equation}
Each distribution function depends on four constants $A_i, a_i, d_i, E_i$ (with $i=1,2$), so that in total the solution for the self-consistent potential $\Phi$ from the Poisson equation
\begin{equation} \label{PE2C}
\nabla^2 \Phi (r)= 4\pi G\left(\int f^{(\nu)}_{T,1}\ d^3v + \int f^{(\nu)}_{T,2}\ d^3v\right)
\end{equation}
requires a study with eight arbitrary constants. In practice, from the point of view of dimensionless parameters, by means of physical arguments we will reduce our investigation to a two-parameter space; of
course, if desired, we could loosen some of the physical constraints that we are
going to impose and thus extend our discussion.

As noted in the Introduction, different physical arguments can motivate the study of two-component models. Here we focus on the case in which we distinguish one population of lighter stars (let $m_1$ be the representative mass of its individual stars and $M_1$ its associated total mass) from a second population of heavier stars (with $m_2>m_1$ and in general, $M_2\neq M_1$), so that the total mass of the cluster is $M=M_1 + M_2$. As for the one-component case, we rescale the problem to a dimensionless form, by referring  to a length scale and to an energy scale based on the constants associated with the lighter component. In particular, we define the dimensionless radius $\xi = r a_1^{1/4} d_1$ and the dimensionless potential $\psi = - a_1(\Phi - E_1)$. After such rescaling, we are left with six independent constants. 
To reduce the number of parameters and thus to work in the simplest mathematical context, we make the following assumptions:

\begin{itemize}

\item[-]{We consider a common truncation radius, that is, we take
\begin{equation}
E_1 = E_2 = E_e~.
\end{equation}
Such assumption is frequently made as a starting point for the construction of multi-mass models \cite[e.g., see][]{dacosta:76}.}

\item[-]{We consider two-component models in which the total masses associated with the two components are in a given ratio $M_1/M_2$. Reasonable values for this ratio are suggested by models of the evolution of stellar populations, as briefly described in Appendix \ref{app:1}. Obviously, this can be seen as a requirement on the ratio of the normalization factors $A_1/A_2$. In practice, for a globally self-consistent model this constraint can be written as 
\begin{equation}
\frac{A_1 a_1^{-3/2}}{A_2 a_2^{-3/2}} \frac{\int_0^{\xi_{tr}}\hat{\rho}_1\xi^2d\xi}{\int_0^{\xi_{tr}}\hat{\rho}_2\xi^2d\xi} = \frac{M_1}{M_2}~; \label{M1M2}
\end{equation}
(for the notation $\hat{\rho}_i$, see Eq. (4)).
For a desired mass ratio, the equation is basically a relation for the constant $A_2 a_2^{-3/2}$ in terms of $A_1 a_1^{-3/2}$, but the precise relation has to be worked out iteratively from the global solution.}

\item[-]{We choose a given value for the single-mass ratio $m_1/m_2$ (reasonable values for this ratio are suggested by stellar-population models, as described in Appendix \ref{app:1}) and impose {\it partial} energy equipartition in the {\it central} regions of the system by means of the dimensionless parameter $\eta = 0.2$ (the definition of $\eta$ is given a few lines below). The way in which equipartition is incorporated is not unique (e.g., see \citealt{kondratev:81}). In its simplest form, as proposed by \citet{dacosta:76}, energy equipartition is sometimes imposed by means of a relation between the energy scales of the form $a_2/a_1= m_2/m_1$. Here we prefer to follow the argument of \citet{miocchi:06}, which recognizes that equipartition is best ensured in the central, more relaxed regions. On the other hand, given the support of recent observations \cite[see][]{bellini:14} and simulations \cite[see][]{trenti:13}, it may be wiser to refer to only partial equipartition, by imposing
\begin{equation}
\left[\frac{a_2}{a_1}\frac{\gamma\left(5/2,\Psi\right)\gamma\left(3/2,a_2\Psi/a_1 \right)}{\gamma\left(3/2,\Psi\right)\gamma\left(5/2,a_2\Psi/a_1 \right)}\right]^{1/2}= \left(\frac{m_1}{m_2}\right)^{-\eta}~.\label{eta}
\end{equation}
The left-hand side of the above equation represents the ratio $\sigma_1(0)/\sigma_2(0)$ of the central velocity dispersions for the two-component model.\footnote{$\gamma$ is the incomplete gamma function defined by $\gamma(s,x)=\int_0^x t^{s-1}e^{-t}~dt$.} Note that at $r = 0$ the one-component distribution function is trivial, because the dependence on $J$ drops out and $\Phi = \Phi(0)$, so that Eq.~\eqref{eta} is expressed in closed form in terms of the relevant constants and of the concentration parameter $\Psi = -a_1[\Phi(0) - E_e]$. Full equipartition is marked by $\eta = 1/2$; from their simulations, also in view of an argument by \citet{spitzer:69}, \citet{trenti:13} suggest $\eta = 0.2$ for specific cases. In the following we will refer to this case of partial equipartition (for a recent investigation on energy equipartition in globular clusters, see also \citealt{bianchini:16})}. 

\item[-]{We assume that the radial scales that define the size of the radially biased anisotropic outer envelope are the same for the two components, that is
\begin{equation}
d_2 a_2^{1/4} =  d_1 a_1^{1/4}~.
\end{equation}
This is only a qualitative argument, meant to recognize that one of the possible causes of radially-biased pressure anisotropy is incomplete violent relaxation, which is a collisionless relaxation process that acts in the same way on stars of different masses \cite[see also][]{gunn:79}. For convenience in the numerical
calculation of the models, we decided to adopt the radial scale $d a^{1/4}$ as a
proxy for the radius of transition from isotropic core to anisotropic envelope;
by inspecting one-component and two-component models, we confirm that indeed
this scale identifies approximately the anisotropy radius $r_{\alpha}$.}

\end{itemize}

To summarize, our two-component models depend on eight constants. In practice, by taking a common truncation radius and a common pressure anisotropy scale for the two components and by fixing the values of the ratios $M_1/M_2$, $m_1/m_2$ (and of $\eta$), the relations introduced above reduce the number of free constants to four. Two of them are used to rescale the Poisson equation to a dimensionless form, the remaining two define two independent dimensionless parameters, so that the parameter space explored by the family of two-component models considered in the present study is two-dimensional. As in the one-component models, we use as independent structural parameters the central dimensionless potential $\Psi = -a_1[\Phi(0) - E_e]$ and the parameter $\gamma=a_1d_1^2/(4\pi G A_1)$.

\subsection{Mass segregation}
\label{subsec:3.1}

The third condition imposed in the construction of two-component models is meant to incorporate the role of collisions in establishing some sort of equipartition. It is well known that this effect should be accompanied by mass segregation, that is, by a general trend of the lighter component to exhibit a more diffuse distribution with respect to the heavier component. In particular, we note that for our models the central density ratio is given by
\begin{equation}
\frac{\rho_1(0)}{\rho_2(0)} =\frac{A_1}{A_2}\left(\frac{a_2}{a_1}\right)^{3/2} \frac{e^\Psi\gamma\left(3/2,\Psi\right)}{e^{a_2\Psi/a_1 }\gamma\left(3/2,a_2\Psi/a_1 \right)}~,\label{segr}
\end{equation}
which, under the conditions listed in the previous subsection, would be expected to fall below unity from a simple picture of mass segregation (in which the central parts should be dominated by the heavier component).

\begin{figure} 
\resizebox{\hsize}{!}{\includegraphics{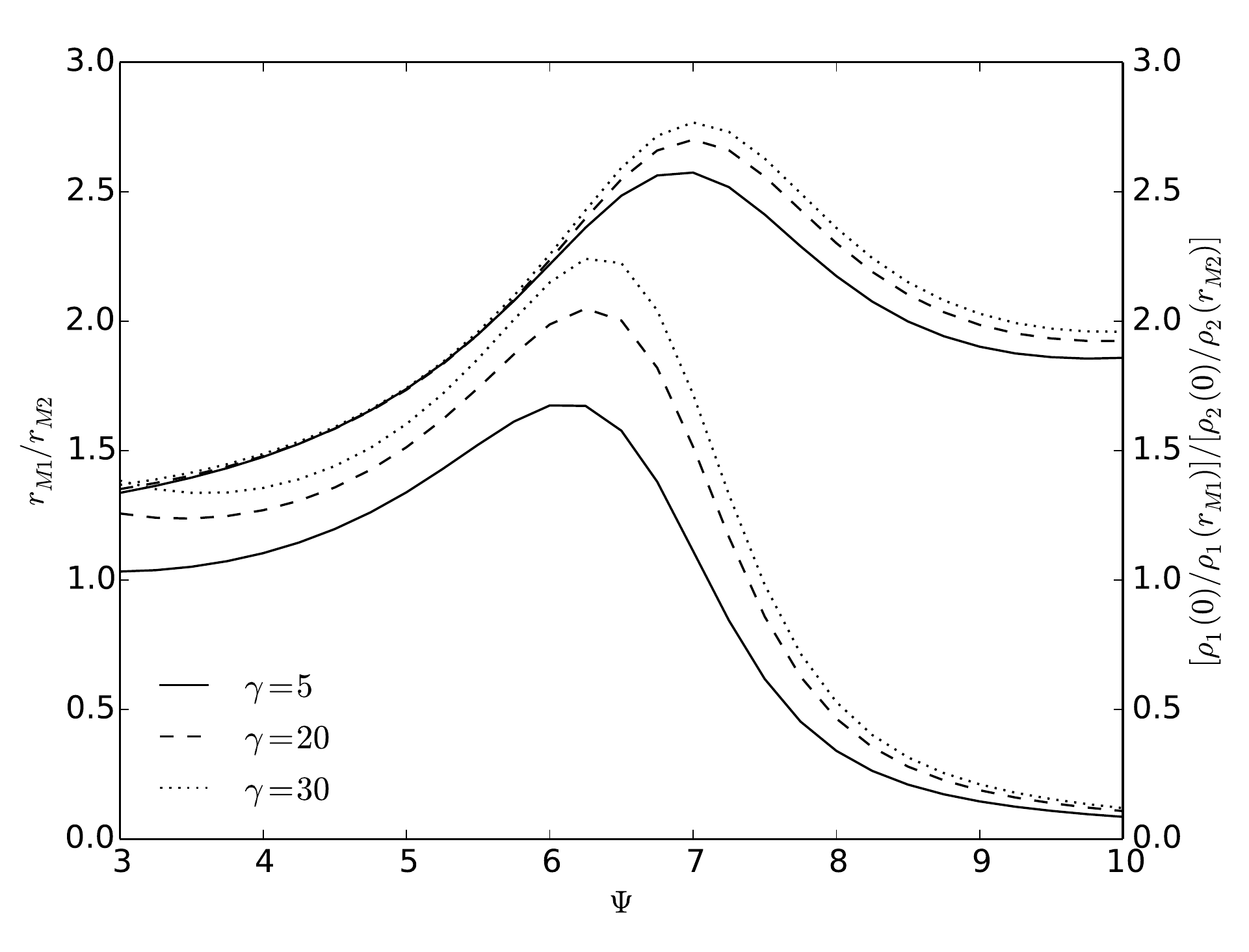}} 
\caption{Relative concentration of the two components as a function of $\Psi$, for selected values of $\gamma$. The upper set of curves represents the ratio $r_{M1}/r_{M2}$ of the half-mass radius of the lighter component to the half-mass radius of the heavier component. The lower set represents the ratio of the density-contrast parameters.}
\label{fig:conc2C}
\end{figure}

As we noted in Subsect.~\ref{subsec:2.2}, when we introduced the concentration parameter $\Psi$ for the one-component models, there are several ways of describing the concentration of a given density profile. Here, we illustrate the result of different definitions that may be adopted. In Fig.~\ref{fig:conc2C} we plot the ratio $r_{M1}/r_{M2}$ of the half-mass radii of the two components and the ratio of the quantities associated with the parameter illustrated in Fig.~\ref{fig:concentration}, that is, of the density contrast of the lighter component $\rho_1(0)/\rho_1(r_{M1})$ to that of the heavier component $\rho_2(0)/\rho_2(r_{M2})$, as a function of $\Psi$, for selected values of $\gamma$. The ratio $r_{M1}/r_{M2}$ exceeds unity for all the models considered and thus it is the more natural parameter to be used to describe the relative concentration of the two components. 

\begin{figure*}
\centerline{
\includegraphics[width=0.48\textwidth]{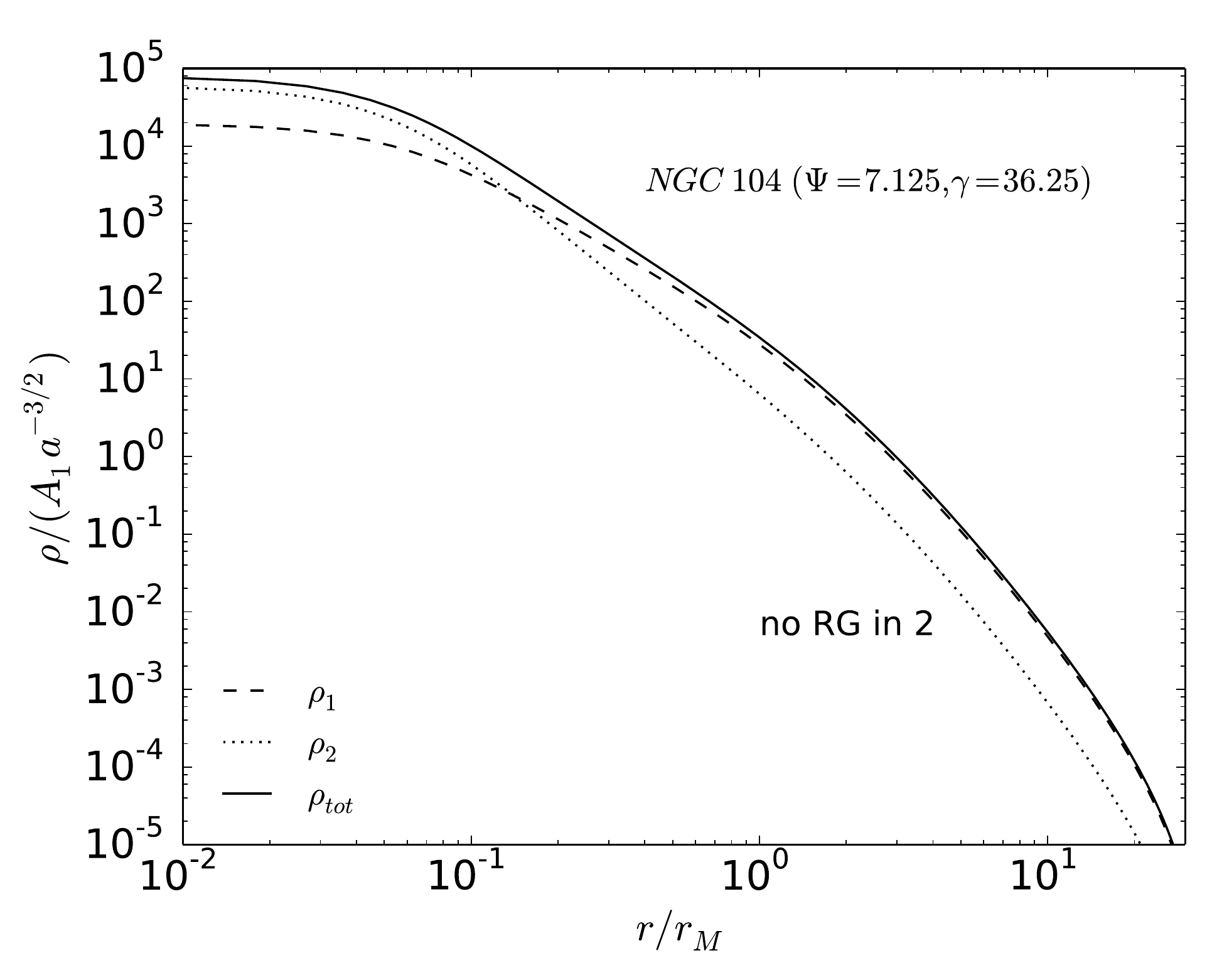}
\includegraphics[width=0.48\textwidth]{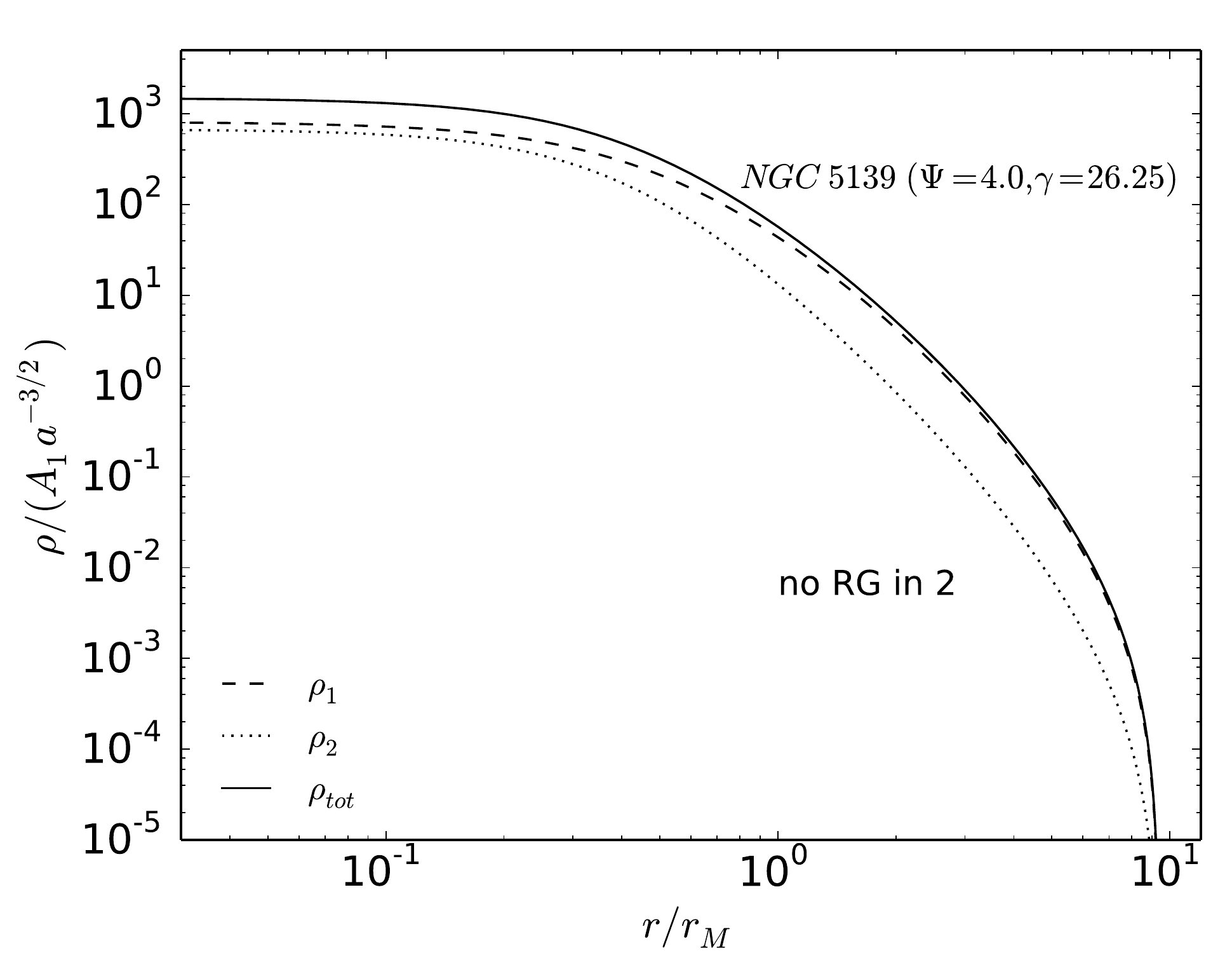}
}
\caption{The left frame shows the density profiles of each component and the total density profile for the best-fit model of 47 Tuc (NGC 104), obtained by the procedure in which RG stars are not included in the heavy component (see text); the right frame shows the corresponding density profiles for the best-fit model of $\omega$ Cen (NGC 5139).}
\label{fig:rho_Om_Tuc}
\end{figure*}

In order to highlight how different types of mass segregation can result from the condition of partial energy equipartition imposed on our models, we report the cases of two selected globular clusters: 47 Tuc and $\omega$ Cen. We have found the two-component dynamical models that best reproduce the observed photometric and kinematic profiles of the two clusters. In Fig.~\ref{fig:rho_Om_Tuc} we plot the density profiles of the two best-fit models found by the procedure in which Red Giant stars are not included among the heavy stars (for a discussion of this fitting procedure, see the next section). The best-fit model of 47 Tuc is characterized by a density profile with a larger density of heavy stars in the central regions. Indeed, this is the type of mass segregation traditionally associated with the tendency of the system to establish energy equipartition. The model of $\omega$ Cen exhibits a qualitatively different mass distribution.

In the next section, devoted to setting the correspondence between dynamical models and observations, we briefly describe how mass segregation has a counterpart in the gradient of the profile of the cumulative mass-to-light ratio, defined as the total mass-to-light ratio for a sphere of given radius $r$.


\section{Fitting the data with dynamical models}
\label{sec:4}
We have performed a combined photometric and kinematic fit to the data available for a set of globular clusters, following a procedure very similar to that used in \citetalias{zocchi:12}. In the present analysis we have decided to minimize a combined chi-square function, which is defined as the sum of the photometric and the kinematic contributions. Differently from the fits reported in \citetalias{zocchi:12} by means of one-component non-truncated $f^{(\nu)}$ models, the fits presented here, based on the $f_T^{(\nu)}$ models, are characterized by one additional parameter ($\gamma$) strictly connected with the truncation.

\subsection{The issue of the mass-to-light ratios}
\label{subsec:4.1}
\subsubsection{Mass-to-light ratios for one-component models}

In the application of one-component models, we follow the general assumption that a constant mass-to-light ratio adequately describes the stellar population, imagined to be homogeneous. This assumption allows us to convert projected mass densities $\Sigma(R)$ into surface luminosity densities $l(R)$ by means of a simple relation of proportionality. Then, the mass-to-light ratio is found as one of the parameters determined by the fit (see Appendix B of \citetalias{zocchi:12}).

\subsubsection{Mass-to-light ratios for two-component models}
In general, for the two-component models we consider the surface luminosity profile as the sum of two contributions:
\begin{equation}
\label{constML}
l(R)=\Sigma_1(R)\left(\frac{M}{L}\right)_1^{-1}+\Sigma_2(R)\left(\frac{M}{L}\right)_2^{-1}~.
\end{equation}
Then, we have performed two different types of fit:
\begin{enumerate}[label=(\roman*)]
\item {In the first procedure, we consider the heavier component made of only dark remnants. Therefore, the fit is similar to that for elliptical galaxies in the presence of a dark matter component. In other words, the photometric fit is carried out by omitting the $\Sigma_2$-term in Eq.~(\ref{constML}). Then the kinematic fit is performed by considering only the velocity dispersion profile relative to the {\it lighter} component, which is the only component assumed to be visible.}
\item {In the second type of fit, we include the Red Giant stars (RGs) in the group of the heavier stars (see Appendix \ref{app:1}). In this case, in the photometric fit both components contribute to the surface brightness. Thus, we have explored two possible options: either (a) to assign a reasonable value for the ratio $(M/L)_1/(M/L)_2$, based on the fraction of luminosity expected to come from the RGs and the main-sequence stars present in the system; or (b) to leave the mass-to-light ratio of the heavier component to be determined as a parameter of the best-fit model, and thus to make a prediction on the number of RGs contained in the system. In this paper we report only the results given by option (a), as the best-fit models found with the other option tend to underestimate the contribution of RGs present in globular clusters.\footnote{Typically, RGs are estimated to provide $\approx 60\%$ of the total V-band luminosity and $\approx 0.5\%$ of the total mass of a globular cluster; these values have been computed by evolving a set of stars with masses distributed according to the Kroupa IMF by means of the SSE package \citep{hurley:00}} In this procedure the kinematic fit considers the {\it heavier} component as the kinematic tracer, because most kinematic data come from spectroscopic observations of RGs (i.e., the line-of-sight velocities of RG stars are usually those that are detected for the construction of the observed velocity dispersion profiles).}
\end{enumerate}

\begin{figure*}
\centerline{
\includegraphics[width=0.48\textwidth]{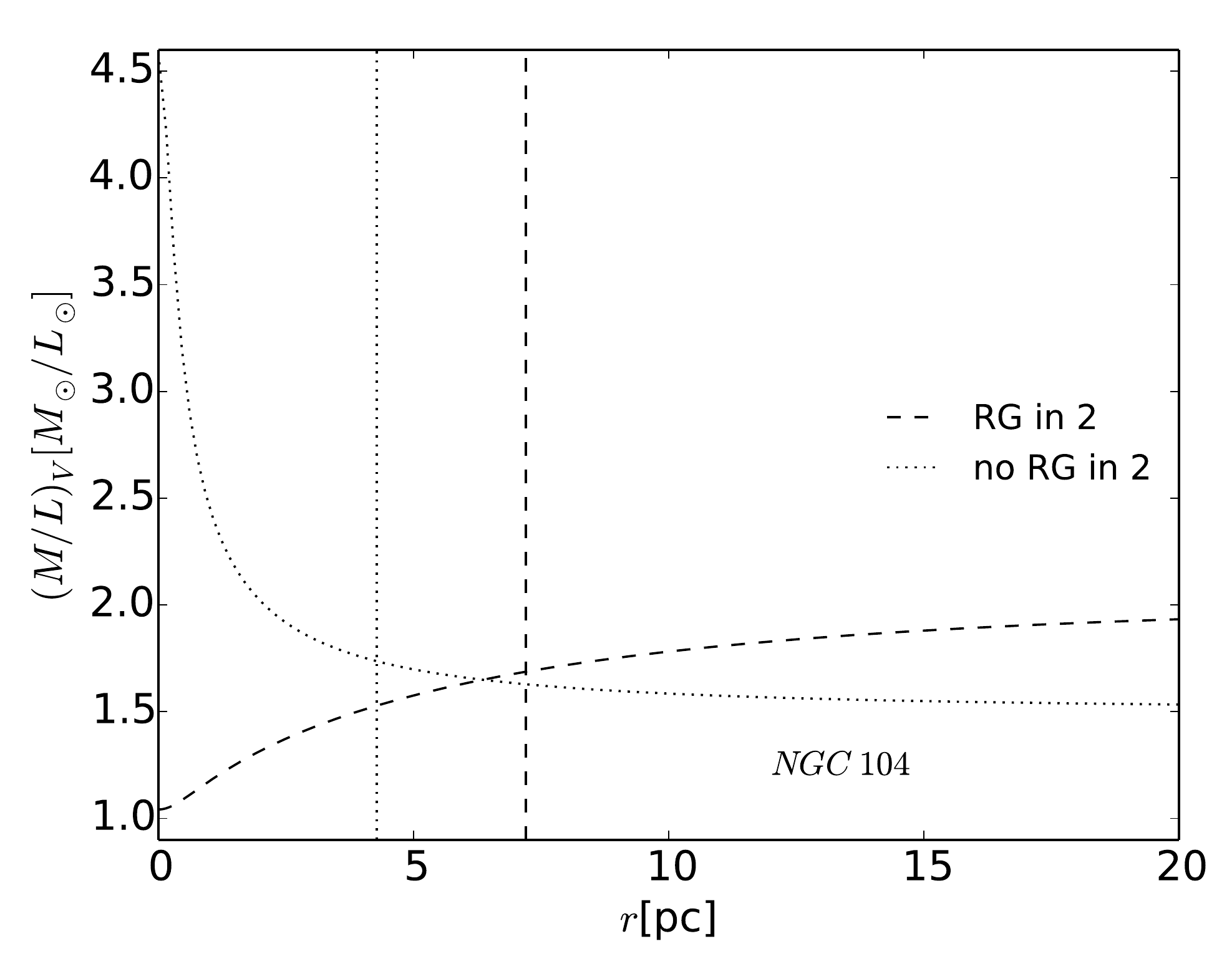}
\includegraphics[width=0.48\textwidth]{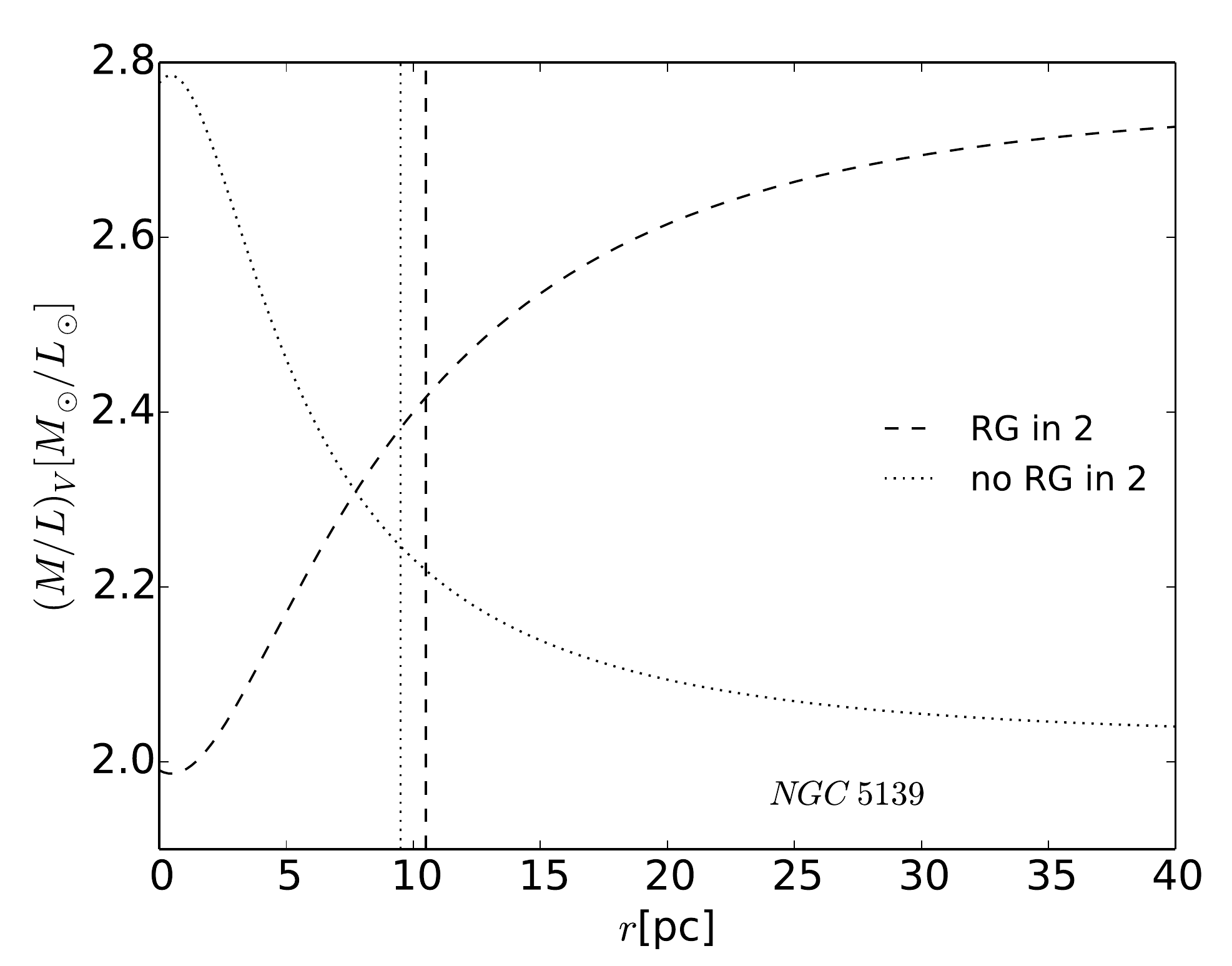}
}
\caption{The cumulative mass-to-light ratio as a function of the intrinsic radius $r$ for the best-fit models of two globular clusters. The best-fit models are found by means of two different procedures, that is,  by taking the heavier component as made of only dark remnants or by including in the heavier component the presence of Red Giants. The vertical lines indicate the position of the total half-mass radius.}
\label{fig:ML}
\end{figure*}

Note that, for the two-component models, the conversion from density profiles to luminosity profiles is not straightforward as in the one-component case, because it depends on the structural characteristics of the system. In particular, it reflects the interconnection between mass segregation and the gradients of mass-to-light ratios. In Fig.~\ref{fig:ML}, we plot the cumulative mass-to-light ratio for two selected globular clusters in their central regions; the behavior of this quantity as a function of the intrinsic radius $r$ changes according to the type of fit considered. On the one hand, in the case in which RGs are not included in the heavier component, the ratio $M/L$ {\it decreases} with $r$ (for the more relaxed cluster 47 Tuc, this trend is more evident). On the other hand, the case in which RGs are included in the heavier component (and in the fitting procedure) is characterized by a mild {\it increase} of the cumulative mass-to-light ratio. For the former case we recover a behavior of the cumulative mass-to-light ratio profile similar to that found by \cite{bosch:06} for the globular cluster M15 (NGC 7078); they suggest that the gradient of the ratio M/L at small radii is likely to be due to the presence of a centrally concentrated population of dark remnants, an interpretation that is also suited to describe the result of our fit. 

We wish to emphasize that in this paper we are not aiming at providing improved dynamical models for selected clusters. Rather, we wish to demonstrate, by means of the mathematically simplest framework, how different ways of using a multi-component dynamical model actually lead to different pictures of the internal structure of globular clusters, especially in relation to mass segregation and gradients of mass-to-light ratios.

\subsection{Fits with one-component models}
\label{subsec:4.2}
The data sets considered in this paper are the same as used by \citetalias{zocchi:12}. 
\begin{table}
\caption{Selected sample of globular clusters.}
\centering
\begin{tabular} {lcccccc}
\toprule			
\toprule
Globular cluster				&$d_\sun$		&$\log{T_c}$	&$\log{T_M}$	&$N_p$		&$N_k$\\
\midrule
NGC 362					&8.6			&7.76		&8.93		&239			&8\\
NGC 7078 (M15)			&10.4		&7.84		&9.32		&310			&35\\
NGC 104 (47 Tuc)			&4.5			&7.84		&9.55		&231			&16\\
NGC 6121 (M4)				&2.2			&7.90		&8.93		&228			&10\\
NGC 6341 (M92)			&8.3			&7.96		&9.02		&118			&8\\
\midrule
NGC 6218 (M12)			&4.8			&8.19		&8.87		&143			&11\\
NGC 6254 (M10)			&4.4			&8.21		&8.90		&162			&6\\
NGC 6656 (M22)			&3.2			&8.53		&9.23		&143			&7\\
NGC 3201					&4.9			&8.61		&9.27		&80			&16\\
NGC 6809 (M55)			&5.4			&8.90		&9.29		&114			&13\\
NGC 288					&8.9			&8.99		&9.32		&85			&6\\
\midrule
NGC 5139 ($\omega$ Cen)		&5.2			&9.60		&10.09		&72			&37\\
NGC 2419					&82.6		&9.87		&10.63		&137			&6\\
\bottomrule
\end{tabular}
\tablefoot{For each globular cluster the following quantities are recorded: distance from the Sun (kpc); logarithm of the core relaxation time (years); logarithm of the half-mass relaxation time (years); number of points in the surface brightness profile; and number of points in the  velocity dispersion profile (adapted from \citetalias{zocchi:12}).}
\label{tab:Sam_GCs}
\end{table}
For convenience, in Table \ref{tab:Sam_GCs} we report some distinctive quantities for the sample of 13 Galactic GCs selected for this paper.

\begin{figure*}
\centering
\includegraphics[width=0.85\textwidth]{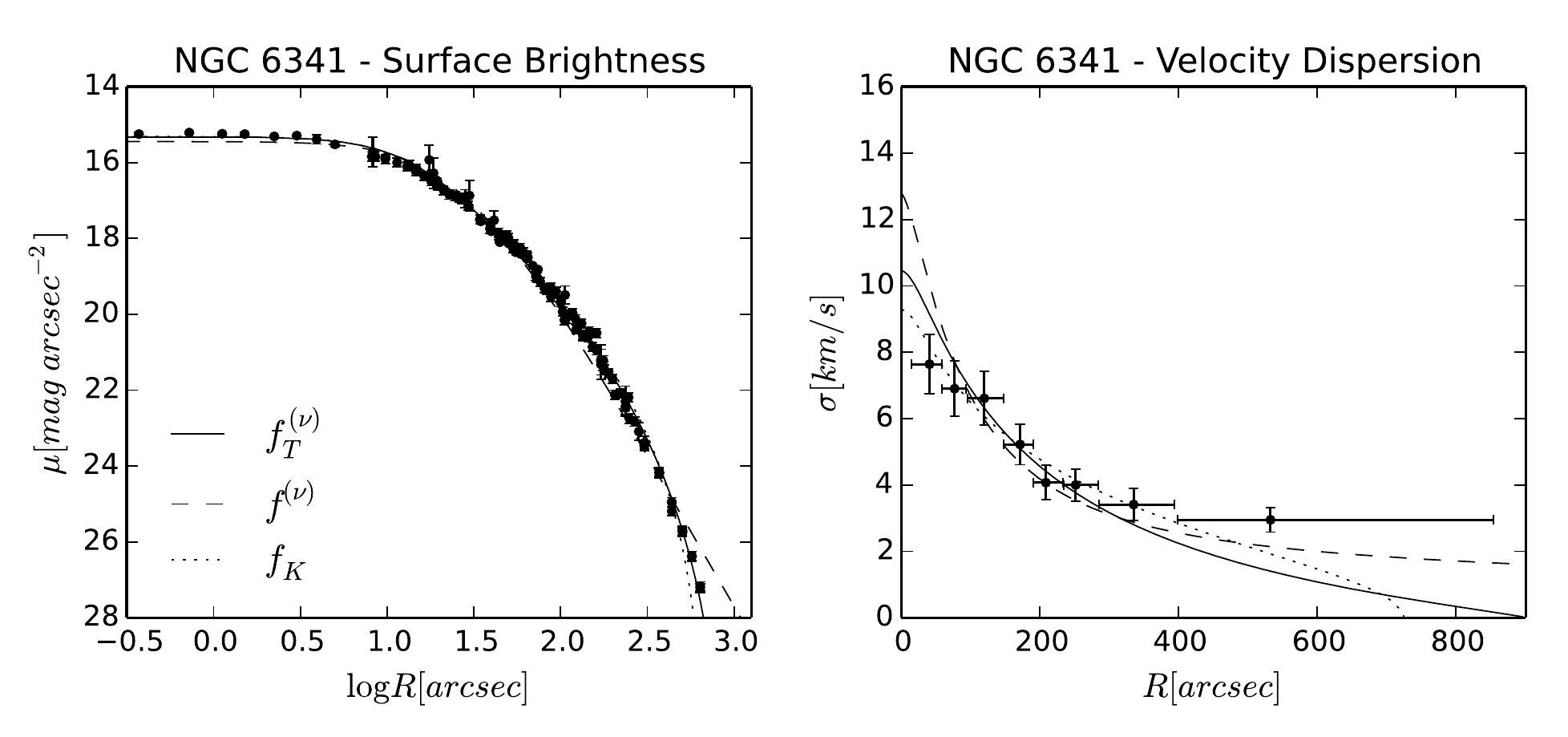}
\includegraphics[width=0.85\textwidth]{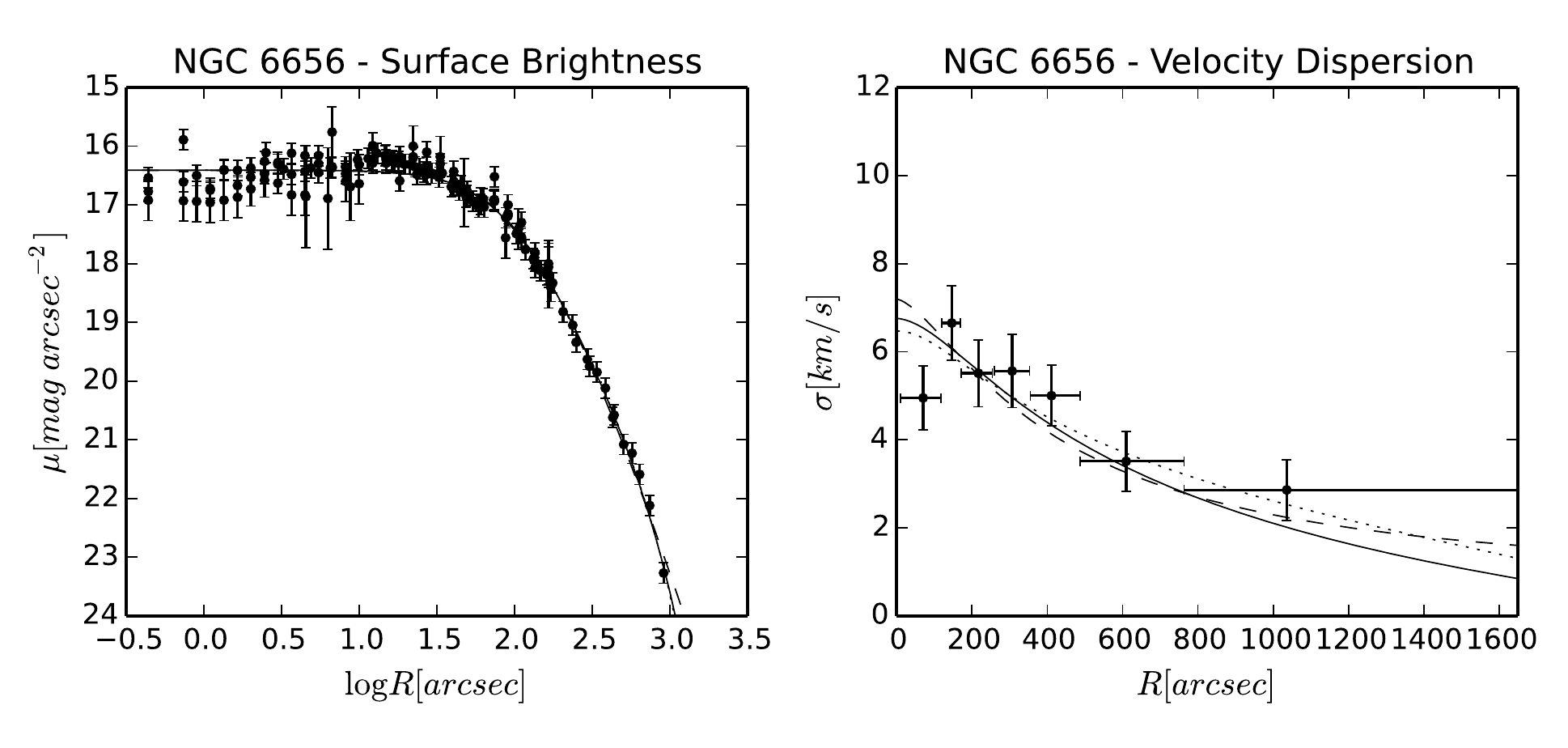}
 \includegraphics[width=0.85\textwidth]{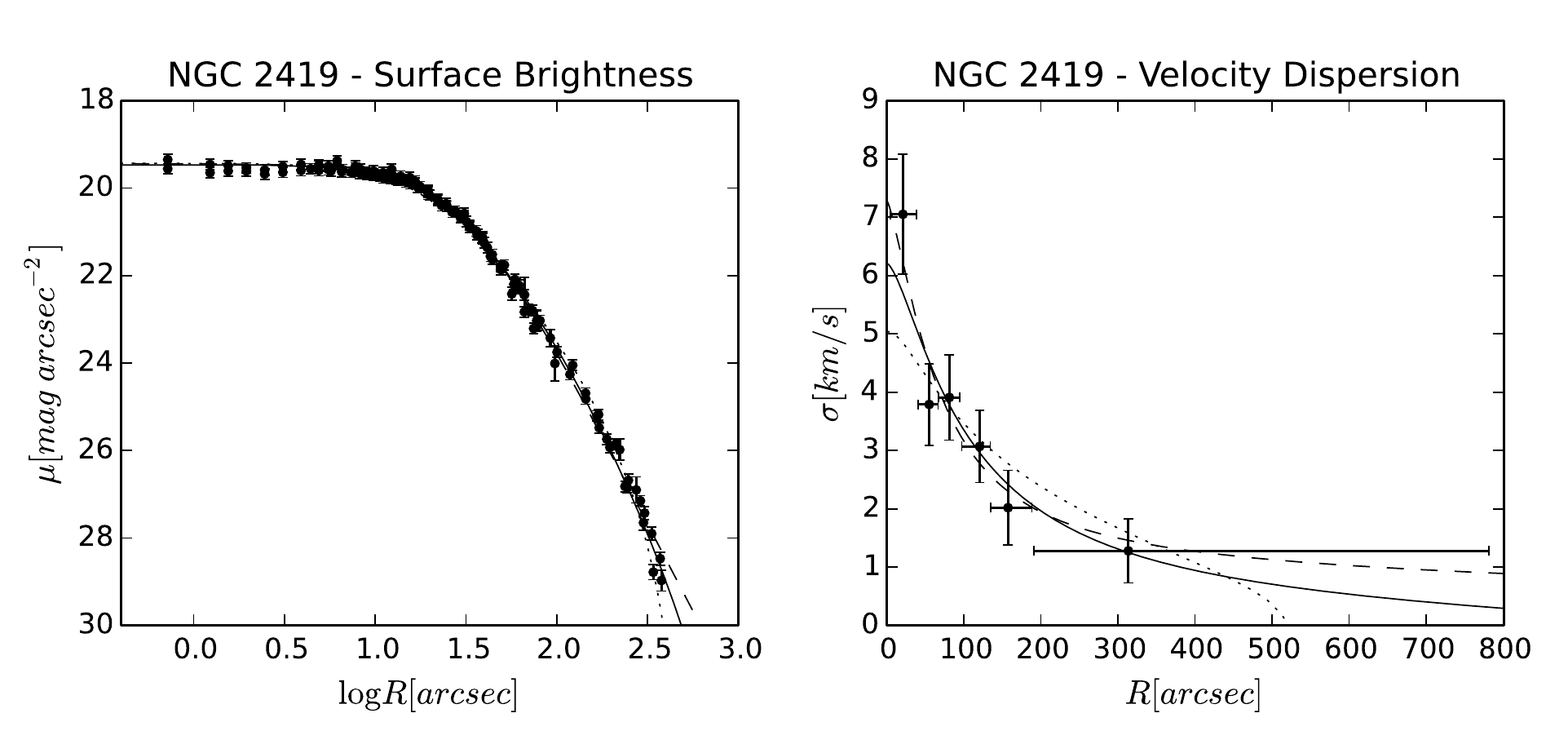}
\caption{Photometric and kinematic fits for three globular clusters of the sample. Each cluster is representative of its relaxation class as identified by the core relaxation time $T_c$ (for NGC 6341, $\log{T_c}\approx7.96$; for NGC 6656, $\log{T_c}\approx8.53$; for NGC 2419 $\log{T_c}\approx9.87$). The curves represent the surface brightness profile (left panels) and the velocity dispersion profile (right panels) calculated by means of dynamical models. In particular, dotted lines correspond to King models; dashed lines to the non-truncated $f^{(\nu)}$ models, and solid lines to the $f^{(\nu)}_T$ models. In all panels, the dots are the observed data. For each data-point, errors are shown as vertical bars; in the case of the velocity dispersion profile, the horizontal bars indicate the size of the the radial bin used to calculate each data point. The King profiles, the  $f^{(\nu)}$ profiles, and the observed data are taken from \citetalias{zocchi:12}.}
\label{fig:aa}
\end{figure*}

In Fig.~\ref{fig:aa} we show the best-fit surface brightness and velocity
dispersion profiles for 3 of the selected GCs, which are displayed in order of
increasing core relaxation time. The dimensionless parameters of the fits and
the values of the reduced chi-squared are listed in Table~\ref{tab:fits_par}.
For the statistical analysis we have followed the procedure used by
\citetalias{zocchi:12}. From an inspection of the way the best-fit models are identified, we note that the present models are characterized by significant
degeneracy in parameter space: this is a natural consequence of the introduction
of the additional parameter related to the truncation.

In general, the photometric fits by the $f^{(\nu)}_T$ models are more
satisfactory than those performed by means of the King and $f^{(\nu)}$ models,
for every relaxation class considered (for a comparison of the values of the
reduced chi-squared, see Table~4 in \citetalias{zocchi:12}); indeed, for the
majority of the clusters, the minimum chi-squared is inside the $90 \%$
confidence interval. The improvement with respect to the King and the $f^{(\nu)}
$ models is mainly related to the outer regions of the system, where the
truncation of our models accommodates well the observed brightness profiles.

In addition, the general trends found by \citetalias{zocchi:12} for the non-truncated models are not affected by the truncation significantly. In particular, our models remain able to reproduce the central peak in the velocity dispersion profiles that is characteristic of the least relaxed clusters in the sample (NGC 2419 and NGC 5139).

\begin{table}
\caption{Best-fit parameters for the one-component models.}
\centering
\begin{tabular} {lccccc}	
\toprule			
\toprule
NGC		&$\Psi$ 			&$\gamma$		&$\tilde{\chi}^2_{ph}$		&$\tilde{\chi}^2_{k}$	\\
(1)		&(2)				&(3)				&(4)					&(5)				\\
\midrule
104		&$8.59\pm0.01$		&$19.2\pm0.5$		&1.14				&11.33	\\
288		&$4.76\pm0.13$		&$4.52\pm0.17$		&1.26				&0.89	\\
362		&$7.32\pm0.03$		&$47.2\pm1.6$		&1.19				&2.39	\\
2419		&$5.55\pm0.06$		&$58\pm2$		&1.15				&0.54	\\
3201		&$5.61\pm0.17$		&$31.7\pm5$		&1.14				&2.74	\\
5139		&$4.81\pm0.08$		&$27.6\pm1.6$		&3.05				&2.45	\\
6121		&$7.38\pm0.07$		&$4.07\pm0.2$		&1.35				&0.54	\\
6218		&$5.60\pm0.09$		&$18.0\pm1.1$		&1.12				&0.68	\\
6254		&$5.62\pm0.9$		&$46\pm1.6$		&4.69				&0.60	\\
6341		&$7.41\pm0.02$		&$18.2\pm0.8$		&6.43				&2.96	\\
6656		&$6.37\pm0.13$		&$12.7\pm4$		&1.03				&1.18	\\
6809		&$3.75\pm0.09$		&$8.0\pm0.23$		&1.15				&2.00	\\
7078		&$8.43\pm0.01$		&$46.6\pm0.25$		&3.72				&2.07	\\
\bottomrule
\end{tabular}
\tablefoot{For each cluster, in Cols. (2) and (3) we provide the best-fit parameters that define the dynamical models, together with their formal errors. We then list the values of the photometric reduced chi-square $\tilde{\chi}^2_{ph}$ (Col. 4) and the kinematic reduced chi-square $\tilde{\chi}^2_{k}$  (Col.5).}
\label{tab:fits_par}
\end{table}

In Table \ref{tab:fits_der_prop} we report the values of the truncation radius $r_{tr}$, the projected core radius $R_c$ (that is the radial location where the surface brightness equals half its central value), and the intrinsic half-mass radius $r_M$. Then we list other relevant quantities, in particular, the total mass $M$, the central density $\rho_0$, and the V-band mass-to-light ratio $(M/L)_V$. For our anisotropic models we have also calculated the intrinsic anisotropy radius $r_\alpha$ defined as the radius where $\alpha(r_\alpha)=1$ and the global anisotropy parameter $\kappa$ (see Subsect. \ref{subsec:2.4}).

\subsubsection{A comparison with the King models}
\label{subsubsec:4.2.1}
No systematic trends are found. The only exception is represented by the truncation radius, which is generally larger for the $f_T^{(\nu)}$ models, in line with the general finding that the photometric profiles appear to possess a smoother truncation than that of King models (see McLaughlin \& van der Marel 2005).

\subsubsection{Radial-orbit instability}
\label{subsubsec:4.2.2}
One of the points noted in the analysis by \citetalias{zocchi:12} is a general concern about the possible occurrence of the \emph{radial-orbit instability}. \cite{polyachenko:81} argued that this instability would occur when the anisotropy parameter $\kappa=2K_r/K_T$, the ratio of the radial contribution to the tangential contribution to the total kinetic energy, exceeds $1.7\pm0.25$. 

In this respect, for some of the globular clusters considered by \citetalias{zocchi:12} (e.g., NGC 6254) the non-truncated $f^{(\nu)}$ models might not be applicable. The truncation in our $f_T^{(\nu)}$ models tends to reduce the global value of the radial contribution to the kinetic energy (see Fig.~\ref{fig:kappa}), bringing $\kappa$ down to values typically associated with stability. Of course, a test by N-body simulations would be desired to confirm this point, but obviously this would bring us well beyond the goals of the present paper.

\begin{table*}
\caption{Derived physical parameters from the best-fit one-component models.}
\centering
\begin{tabular} {l|cc|cc|cc|cc|cc|cc|c|c}	
\toprule
NGC 	&\multicolumn{2}{c|}{$r_{tr}$ }		&\multicolumn{2}{c|}{$R_c$	}		&\multicolumn{2}{c|}{$r_M$\tablefootmark{*}}		&\multicolumn{2}{c|}{M}	&\multicolumn{2}{c|}{$\log{\rho_0}\ $}	&\multicolumn{2}{c|}{$(M/L)_V$}	&$r_\alpha$  &$\kappa$\\

\midrule
104		&2336	&3641	&22.6	&22.09		&5.60	&5.22	&7.18	&7.77	&5.01	&5.09	&1.34	&1.63	&13.98	&1.20\\			
288		&896		&835		&79.42	&77.32		&7.50	&7.53	&0.74	&0.76	&2.04	&2.09	&1.88	&2.20	&23.82	&1.61\\			
362		&897		&1600	&9.88	&10.75		&2.65	&2.26	&1.87	&1.81	&4.83	&4.78	&1.05	&1.09	&5.35	&1.26\\		
2419		&517		&1163	&18.58	&20.20		&26.17	&23.89	&7.84	&9.50		&1.87	&1.90	&1.72	&2.17	&42.85	&1.38\\		
3201		&1533	&2278	&71.58	&74.77		&5.12	&5.03	&1.31	&1.10	&3.01	&2.98	&1.91	&1.99	&11.29	&1.28\\	
5139		&2861	&3549	&127.68	&163.53		&10.02	&10.24	&26.45	&31.16	&3.54	&3.39	&1.93	&2.87	&19.83	&1.34\\
6121		&3144	&2555	&71.31	&69.49		&3.69	&3.72	&0.65	&0.66	&3.66	&3.68	&1.10	&1.20	&17.52	&1.10\\
6218		&982		&1105	&47.81	&50.26		&3.29	&3.22	&0.61	&0.66	&3.31	&3.31	&1.96	&1.50	&8.57	&1.22\\
6254		&1126	&2191	&50.02	&51.16		&3.31	&3.18	&1.53	&1.72	&3.74	&3.80	&1.61	&2.09	&6.32	&1.33\\
6341		&724		&900		&14.18	&14.42		&3.12	&2.96	&2.86	&3.39	&4.63	&4.71	&1.83	&2.15	&9.51	&1.18\\
6656		&2057	&2224	&80.92	&80.74		&4.18	&4.17	&2.08	&2.14	&3.64	&3.65	&1.11	&1.11	&13.95	&1.16\\	
6809		&1072	&1084	&110.09	&109.23		&5.90	&5.89	&0.60	&0.60	&2.21	&2.24	&1.12	&1.14	&12.55	&1.29\\
7078		&560		&4289	&7.51	&5.46		&2.97	&2.88  	&3.98      	&3.95	&5.21	&5.54	&1.12	&1.22	&5.62	&1.28\\
\bottomrule
\end{tabular}
\tablefoot{For each cluster listed in the first column, in double-column form we provide the relevant physical quantities derived from the King models (as reported in \citetalias{zocchi:12} - left columns) and from our truncated anisotropic models $f_T^{(\nu)}$ (right columns). In single-column form, as last items, we provide the anisotropy radius for the best-fit $f_T^{(\nu)}$ models and the global anisotropy parameter $\kappa$. The truncation radius $r_{tr}$ and the core radius are expressed in units of arcsec; the intrinsic half-mass radius and the anisotropy radius in pc. The total mass is expressed in units of $10^5~M_\sun$ and the central mass density $\rho_0$ in $M_\sun~\text{pc}^-3$. Finally, the mass-to-light ratio is given in solar units $M_\sun/L_\sun$. \\
\tablefoottext{*}{Most values of the half-mass radii for the King models reported in \citetalias{zocchi:12} are incorrect; in the present paper we report the corrected values.}}
\label{tab:fits_der_prop}
\end{table*}

\subsection{Fits with two-component models}
\label{subsec:4.3}
As anticipated in the previous sections, in order to address the issue of mass segregation in the simplest mathematical framework, we have studied the performance of our two-component models in fitting two globular clusters characterized by different relaxation conditions: 47 Tuc (NGC 104) and $\omega$ Cen (NGC 5139).

\begin{figure*}
\centering
\includegraphics[width=0.85\textwidth]{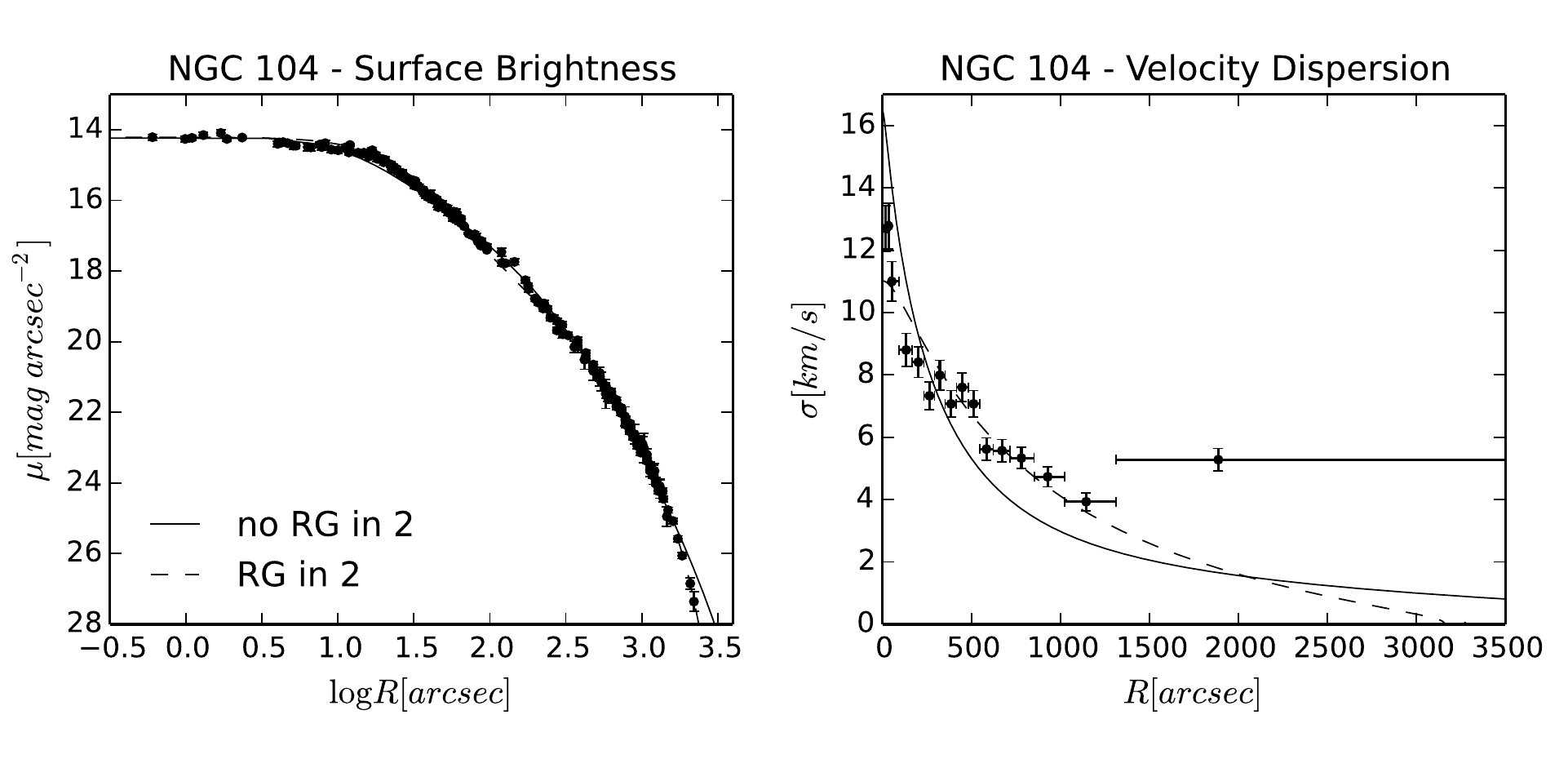}
\includegraphics[width=0.85\textwidth]{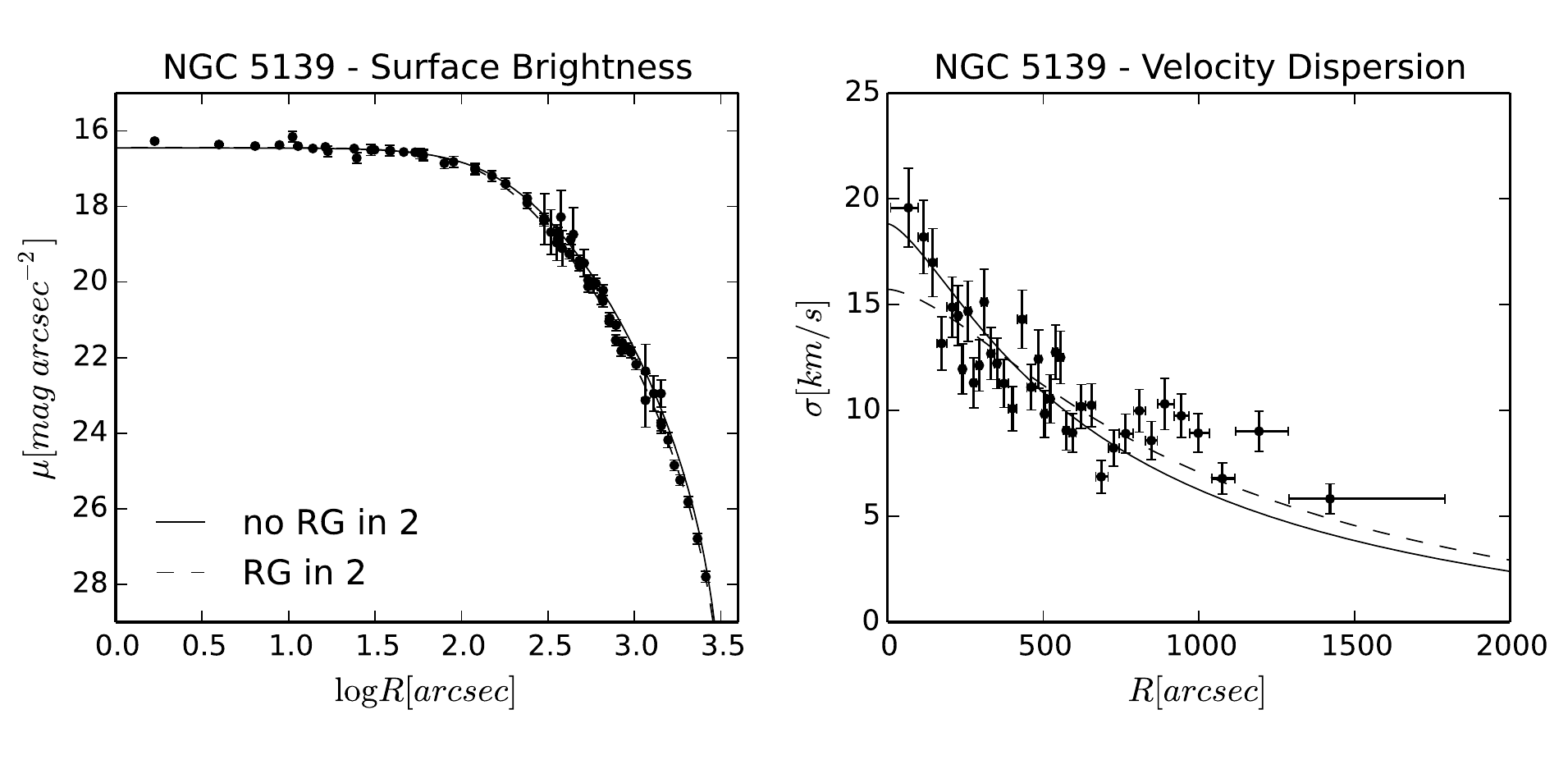}
\caption{Photometric and kinematic fits for NGC 104 and NGC 5139. The curves represent the surface brightness profile (left panels) and the velocity dispersion profile (right panels) calculated by means of two-component models in two ways: by taking the heavier component as made of only dark remnants or by including in the heavier component the presence of Red Giants.}
\label{fig:aa1}
\end{figure*}

The photometric and kinematic fits for these clusters are presented in Fig.~\ref{fig:aa1}. The fits are performed by means of the two procedures outlined in Subsect.~\ref{subsec:4.1}. In particular, for the procedure in which RGs are included in the heavier component, we have assumed that RGs contribute $60\%$ of the total luminosity of the cluster in the V-band.

As in the previous subsection, we report the best-fit parameters (see Table \ref{tab:fits2_par}) and some relevant physical quantities (see Table \ref{tab:fits2_der_prop}). 

\begin{table}
\caption{Dimensionless best-fit parameters for the two-component models.}
\centering
\begin{tabular} {lcccc}	
\toprule			
\toprule
				&\multicolumn{2}{c}{NGC $104$}			&\multicolumn{2}{c}{NGC $5139$} \\
				\cmidrule(lr){2-3}					\cmidrule(lr){4-5} 
				&RG		&no RG					&RG			&no RG \\
\midrule
$\Psi$ 			&$7.00$	&$7.12$					&$4.00$		&$4.00$\\
$\gamma$			&$12.5$	&$36.25$					&$22.50$		&$26.25$\\
$\tilde{\chi}^2_{ph}$	&1.61	&4.89					&1.89		&5.61\\
$\tilde{\chi}^2_{k}$	&8.87	&22.23					&2.26		&3.23\\
\bottomrule
\end{tabular}
\tablefoot{For two clusters considered either by including or by not including RG stars in the heavier component, we provide the best-fit parameters that define the dynamical models ($\Psi$, $\gamma$). We then list the values of the reduced photometric chi-square $\tilde{\chi}^2_{ph}$ and the reduced kinematic chi-square $\tilde{\chi}^2_{k}$.}
\label{tab:fits2_par}
\end{table}

\begin{table}
\caption{Derived physical parameters from the best-fit two-component models.}
\centering
\begin{tabular} {lcccc}		
\toprule	
\toprule
							&\multicolumn{2}{c}{NGC $104$}		&\multicolumn{2}{c}{NGC $5139$} \\
							\cmidrule(lr){2-3}				\cmidrule(lr){4-5} 
							&RG		&no RG				&RG			&no RG \\
				
\midrule
$r_{tr}$					&71		&153					&90			&89\\
$R_{c,1}$				&34.1	&16.3				&176			&155\\
$R_{c,2}$				&21.1	&10.2				&133			&117	\\
$r_{M,1}$ 				&8.7		&5.2					&11.7		&10.4\\
$r_{M,2}$ 				&3.3		&1.8					&7.9			&7.0\\
$M_1$				&7.0		&5.5					&25.6		&22.7\\
$M_2$				&2.3		&1.8					&8.5			&7.5\\
$\log{\rho_{0,1}}$		&4.3		&5.1					&3.19		&3.30\\
$\log{\rho_{0,2}}$		&4.8		&5.6					&3.11		&3.21\\
$(M/L)_{V,1}$			&3.76	&1.13				&5.31		&1.53\\
$(M/L)_{V,2}$			&0.83	&/					&1.18		&/\\
\bottomrule
\end{tabular}
\tablefoot{For two clusters considered either by including or by not including RG stars in the heavier component, we provide the relevant physical quantities relative to the light component 1 and the heavy component 2. The truncation radius $r_{tr}$ and the half-mass radius are expressed in pc; the core radius $R_c$ is expressed in units of arcsec. The total mass is expressed in units of $10^5~M_\sun$ and the central mass density $\rho_0$ in $M_\sun~\text{pc}^-3$. Finally, the mass-to-light ratio is given in solar units $M_\sun/L_\sun$ }
\label{tab:fits2_der_prop}
\end{table}

The two-component models appear to provide good fits to the observed profiles, thus supporting the hypotheses imposed in their construction. For both clusters the fits performed with the procedure that includes RG stars in the heavier component appear to be better. This is particularly evident for the case of 47 Tuc, for which the best-fit model corresponding to the case without RGs in the heavier component does not reproduce the kinematic profile adequately. We then argue that the role of the stars used as kinematic tracer becomes important when we consider more relaxed environments. In turn, the fit to $\omega$ Cen suggests that its stellar population is reasonably homogeneous and mass segregation is probably negligible.

\section{Conclusions and perspectives}
\label{sec:5}

In this paper we have constructed a new class of truncated anisotropic models as an extension of the so-called $f^{(\nu)}$ models, introduced by \citet{stiavelli:87} to describe elliptical galaxies interpreted as the result of incomplete violent relaxation. Such $f_T^{(\nu)}$ models have been applied to perform a combined photometric and kinematic study of a sample of Galactic globular clusters. 

In the first part of the paper, we have constructed one-component truncated models, to describe a stellar system made of a single homogeneous stellar population.  From our analysis, the new class of models is found to be well suited to describe the globular clusters of a sample studied earlier. We have compared our fits with those performed for the same sample of globular clusters by \citetalias{zocchi:12} by means of King  and $f^{(\nu)}$ models.  In general, the new truncated models represent the surface brightness profiles better, especially in the outer parts of the systems.  In addition, the models tend to reproduce the inner parts of the velocity dispersion profiles better than the King models.  As also noted by \citetalias{zocchi:12}, this is probably related to the role played by radially-biased pressure anisotropy in partially relaxed clusters. In the $f^{(\nu)}$ and in the $f_T^{(\nu)}$ models, such radial anisotropy is a signature of the process of incomplete violent relaxation, which may have occurred during the initial stages of the evolution of globular clusters; of course, we should be aware that other mechanisms may be responsible for radially-biased pressure anisotropy. In contrast to some cases found earlier by application of the non-truncated $f^{(\nu)}$ models, the  $f_T^{(\nu)}$ models identified by the fits appear to be stable with respect to the radial-orbit instability.

In the second part of the paper, we have extended our analysis by constructing a family of two-component models, with the aim of characterizing in the simplest way a stellar system made of stars with different masses. In fact, if some collisionality is present, stars of different masses are expected to differ in their dynamical evolution, by exhibiting phenomena associated with equipartition and mass segregation. In particular, we have assumed that the stellar system under consideration is made of only dark remnants and main sequence stars, with the possible inclusion of Red Giant stars. RG stars would naturally belong to the component of heavier stars, but obviously differ from the heavy dark remnants from the point of view of their visibility. This raises an interesting modeling problem, that is, the question of the optimal comparison between the two-component models thus constructed and the available photometric and kinematic data. To explore the relevant underlying modeling issues, the new two-component models have been tested on two globular clusters characterized by different relaxation conditions. They generally provide satisfactory fits to the observed photometric and kinematic profiles, in particular when RGs are included in the fitting procedure, by considering their contribution as heavy stars to the photometric profile and their role in tracing the kinematics of the clusters. Interestingly, from our two-component models only the more relaxed cluster (47 Tuc) exhibits the signature of mass segregation in a prominent way.

The two-component models that we have introduced address the effects induced by collisionality on stars characterized by different masses. This is only one particular application of two-component models. We plan to consider soon the construction of two-component models aimed at addressing the issue of dark matter in globular clusters and of others able to touch on the issue of the recently observed multiple stellar populations \cite[generally thought to represent different episodes of star formation; see][]{gratton:12}.

\begin{acknowledgements}
We would like to thank A.L. Varri and D. Heggie for helpful suggestions and M. Trenti for useful conversations about the topics addressed in this paper. We are also grateful to S. Degl'Innocenti and to P.G. Prada Moroni for discussions related to the properties of stellar populations in globular clusters. AZ thanks M. Gieles and V. H\'enault-Brunet for many interesting discussions. This work was partially supported by the Italian MIUR. AZ acknowledges financial support from the Royal Society (Newton International Fellowship).
\end{acknowledgements}

\nocite{*}
\bibliographystyle{aa} 
\bibliography{AA_2016_28274.bbl} 

\begin{appendix}
\section{Dark remnants, Red Giants, and main-sequence stars}
\label{app:1}
In the present study, we have simplified the discussion of the structure of a system made of stars of different masses by grouping the various stars into two components, light stars of mass $m_1$ (and total associated mass $M_1$) and heavy stars of mass $m_2$ (and total associated mass $M_2$). Real globular clusters are extremely complex, because they contain not only stars with basically a continuous spectrum of masses, but also objects, such as binary stars, that fall outside the paradigm of the equations traditionally used in stellar dynamics. The main goal of this appendix is to determine ``reasonable" estimates for the mass ratios $m_2/m_1$ and $M_2/M_1$ to be used in our idealized models, as introduced in Sect.~\ref{sec:3}.

Most of the objects that are naturally assigned to the heavier component (and collectively should make most of the mass $M_2$) are often called ``dark remnants".  In fact, white dwarfs, neutron stars, and black holes are not expected to contribute much to the surface brightness profile of the cluster. In turn, most of the mass $M_1$ is expected to be made of low-mass (typically below $0.5~M_\sun$) main-sequence stars. There remains a third class of stars, the Red Giant stars, which are expected to belong to the heavier component (because their mass is thought to be in the range $0.7-0.8~M_\sun$, very similar to the average mass of the remnants; see below), with only minor contribution to $M_2$; yet, they are expected to contribute significantly to the observed surface brightness and, importantly, are generally used as kinematic tracers, in the sense that they are the main source of the kinematic data points collected by spectroscopic observations. The modeling of globular clusters addressed in this paper is thus significantly different from that used in the two-component description of elliptical galaxies, for which one component represents the luminous collisionless stellar system and the other component the dark matter halo; still, an element of analogy exists, because in both cases the structural profiles of self-consistent two-component models are generally different from those of one-component models.

To estimate a priori some quantities that define our idealized model, we start from the Initial Mass Function (IMF), which defines the initial distribution of stars with mass. Then we make some very simple assumptions about star evolution to estimate how stars have evolved from their initial condition and thus infer some properties of the present distribution of masses. We refer to three different IMFs. The first has been proposed by \citet{salpeter:55} and is a single power law
\begin{equation}
\xi(m)=Dm^{-2.35},
\end{equation} 
where $D$ is a constant. Thus, the quantity $\xi(m)dm$ is the initial number of stars with mass in the range $(m,m+dm)$. The total mass of stars within the mass range ($m_{min}$,$m_{max}$) is given by a simple integration:
\begin{equation}
M=\int ^{m_{max}}_{m_{min}} m\xi(m)dm.
\end{equation}
The corresponding number of stars is 
\begin{equation}
N=\int ^{m_{max}}_{m_{min}} \xi(m)dm,
\end{equation}
so that a mean value for the single mass is given by $m=M/N$. The other forms of IMF considered are taken from \citet{miller:79} and \citet{kroupa:01}. 

More massive stars evolve more rapidly, leaving the main sequence and becoming remnants after a relatively rapid transition in the giant branch. For our purposes, we assume that the main-sequence stars with masses larger than $\approx 0.8~M_\sun$ become remnants instantly (i.e., in a time very short compared to the age of the cluster). In particular, stars with masses from $0.8$ to $10~M_\sun$ become white dwarfs, those with masses from $10$ to $25~M_\sun$ become neutron stars, and those from $25$ to $100~M_\sun$ end up as black holes (we adopted the same mass ranges used by \citealt{gill:08}). A certain fraction of the initial mass is lost through supernova explosions or gas expelled by planetary nebulae. Thus, for white dwarfs we consider masses in the range $0.5-1.4~M_\sun$; for neutron stars we take masses in the range $1.3-2~M_\sun$, and for the black holes masses in the range $5-10~M_\sun$. Fast evolution is thus assumed to map an initial range of masses $0.8 - 100~M_\sun$ distributed according to the IMF into a present-day mass range $0.5 - 10~M_\sun$ for the remnants. The mass functions of the remnants are thus constructed from the IMF by taking the same slope in the corresponding mass ranges. The number of remnants must be equal to the initial number of the main-sequence stars. Such condition fixes the constant $D$ of the mass function of the remnants. Once the various mass functions have been properly defined, we proceed to calculate the mean mass and the total mass of each group of objects. The results are summarized in Table \ref{tab:main_seq_dar_rem}.

\begin{table}
\caption{Masses of different stellar components.}
\centering
\centering
\begin{tabular} {lccc}
\toprule
\toprule
				&Salpeter			&Kroupa					&Miller-Scalo\\
\midrule
$m_{MS}$			&0.21 $M_{\sun}$	&0.28 $M_\sun$			&0.29 $M_\sun$\\
$m_{WD}$			&0.78 $M_{\sun}$	&0.79 $M_\sun$			&0.77 $M_\sun$\\
$m_{NS}$			&1.59 $M_{\sun}$	&1.59 $M_\sun$			&1.57 $M_\sun$\\
$m_{BH}$			&6.80 $M_{\sun}$	&6.59 $M_\sun$			&6.59 $ M_\sun$\\
$m_{DR}$			&0.85 $M_{\sun}$	&0.81 $M_\sun$			&0.79 $M_\sun$\\
$m_{MS}/m_{DR}$	&0.24			&0.35					&0.37\\
$M_{MS}/M_{DR}$	&3.88			&2.10					&1.47\\
\bottomrule
\end{tabular}
\tablefoot{Mean values for the masses of the typical main sequence star (MS), white dwarf (WD), neutron star (NS) and black hole (BH). By dark remnant (DR) we mean all the possible remnants in the cluster. The last row represents the ratio of the total masses.}
\label {tab:main_seq_dar_rem}
\end{table}

By identifying the light stars of component $1$ with the main-sequence stars and the heavy stars of component $2$ with the remnants, in our models we take,\footnote{Note that this choice violates the Spitzer criterion \cite[see][]{spitzer:69}.}  for simplicity, $m_1=0.25~M_\sun$, $m_2=0.75~M_\sun$, $m_2/m_1=3$ and $M_2/M_1=1/3$.

Therefore, the idealized evolution model considered in this Appendix does not include the presence of RGs in the final state. A posteriori the presence of RGs may be taken into account by considering the mass-to-light ratio of the heavy component in the idealized two-component models as a parameter depending on the number of RGs present in the cluster. By determining the mass-to-light ratio of the heavy component, a fit to the data could thus give an estimate of the number of giants in the cluster. Alternatively, if an estimate of the number of RGs is available independently, we would have an a priori estimate of the mass-to-light ratio for the heavy component and thus test the adequacy of the two-component models under such a constraint.
\end{appendix}

\end {document}